\documentclass[apj]{emulateapj}
\usepackage{amsmath}
\usepackage{graphicx}

\def\lta{\,\raise 0.3 ex\hbox{$ < $}\kern -0.75 em
 \lower 0.7 ex\hbox{$\sim$}\,}
\def\gta{\,\raise 0.3 ex\hbox{$ > $}\kern -0.75 em
 \lower 0.7 ex\hbox{$\sim$}\,}
\newcommand{\be}{\begin{equation}}
\newcommand{\ee}{\end{equation}}

\usepackage{epstopdf}
\usepackage{epsfig}
\usepackage{natbib}
\usepackage{wasysym}
\usepackage{multirow}
\begin{document}

\title{Formation and Structure of Low Density Exo-Neptunes}
\author{Leslie A. Rogers$^{1}$, Peter Bodenheimer$^2$, Jack J. Lissauer$^{3}$,
Sara Seager$^{1,4}$}

\affil{$^1$Department of Physics, Massachusetts Institute of
Technology, Cambridge, MA 02139, USA}

\affil{$^2$UCO/Lick Observatory, Department of Astronomy and Astrophysics, 
University of California, Santa Cruz, CA 95064, USA}

\affil{$^3$Space Science and Astrobiology Division,
MS 245-3, NASA-Ames Research Center, Moffett Field, CA 94035, USA}

\affil{$^4$Department of Earth, Atmospheric, and  Planetary Science, Massachusetts Institute of
Technology, Cambridge, MA 02139, USA}

\begin{abstract}
{\it Kepler} has found hundreds of Neptune-size (2-6 $R_{\oplus}$) planet candidates within 0.5~AU of their stars. The nature of the vast majority of these planets is not known because their masses have not been measured. Using theoretical models of planet formation, evolution and structure, we explore the range of minimum plausible masses for low-density exo-Neptunes. We focus on highly irradiated planets with $T_{eq}\geq500$~K. We consider two separate formation pathways for low-mass planets with voluminous atmospheres of light gases: core nucleated accretion and outgassing of hydrogen from dissociated ices. We show that Neptune-size planets at $T_{eq}=500$~K with masses as small as a few times that of Earth can plausibly be formed core nucleated accretion coupled with subsequent inward migration. We also derive a limiting low-density mass-radius relation for rocky planets with outgassed hydrogen envelopes but no surface water. Rocky planets with outgassed hydrogen envelopes typically have computed radii well below $3~R_{\oplus}$. For both planets with H/He envelopes from core nucleated accretion and planets with outgassed hydrogen envelopes, we employ planet interior models to map the range of planet mass--envelope mass--equilibrium temperature parameter space that is consistent with Neptune-size planet radii. Atmospheric mass loss mediates which corners of this parameter space are populated by actual planets and ultimately governs the minimum plausible mass at a specified transit radius. We find that {\it Kepler}'s 2-6 $R_{\oplus}$ planet candidates at $T_{eq}=500$--1000~K could potentially have masses $\lesssim4~M_{\oplus}$. Although our quantitative results depend on several assumptions, our qualitative finding that warm Neptune-size planets can have masses substantially smaller than those given by interpolating the masses and radii of planets within our Solar System is robust.

\end{abstract}


\section[]{Introduction}
\label{sect:intro}

The first 4.5 months of {\it Kepler} data provide evidence for hundreds of  ``Neptune-size'' (2 -- 6 $R_\oplus$, where $R_\oplus$ is the Earth's radius) planets orbiting within 0.5 AU of their stars \citep{BoruckiEt2011ApJ, BoruckiEt2011AstroPh}. The prevalence of planet candidates within this size range raises questions about both planetary growth and migration of Neptune-size planets. Assuming that many of these candidates are true planets, what are they, how did they form, and why are they so numerous? 
 
 {\it Kepler} measures planetary sizes and orbital periods.  In some cases, planet masses can be estimated from dynamical interactions between the planet and its star (Doppler method) or among planets in a multi-planet system \citep[Transit Timing Variations, TTVs,][]{HolmanEt2010Science}.  However, the masses of most of {\it Kepler}'s Neptune-size planet candidates will be difficult to measure.

We model herein the growth, physical evolution, and interior structure of Neptune-size planets that possess voluminous atmospheres of light gases. 
Our focus is obtaining estimates of the minimum plausible masses of {\it Kepler}'s planet candidates. The maximum plausible mass of a planet of radius  $R_p\lesssim~3R_{\oplus}$ can be estimated from the mass-radius relationship for rocky (Earth-like composition) planets \citep[e.g.,][]{ValenciaEt2006Icarus, SeagerEt2007ApJ, FortneyEt2007ApJ, MarcusEt2010ApJ}. 
 In contrast, estimation of minimum plausible masses requires more detailed modeling of planetary growth, because formation of low-mass planets of solar composition demands complicated and contrived scenarios involving large amounts of mass loss. 
 We consider formation of low-density planets both through core nucleated accretion, and through outgassing of low-molecular weight atmospheres. This work does not consider planet formation via gravitational instability \citep[][]{Boss1997Sci} or tidal downsizing \citep{Nayakshin2010MNRAS, Nayakshin2010bMNRAS, Nayakshin2011MNRAS}.
 
We present new core nucleated planet accretion calculations following the approach of \citet{PollackEt1996Icarus} and \citet{MovshovitzEt2010Icarus}. Whereas all previous papers with this code emphasize the formation of Uranus mass and larger planets, here we present a new application of the code to a lower mass regime $\left(M_p<10~M_{\oplus}\right)$. We push to lower planet masses by modeling formation scenarios where the gas disk dissipates well before rapid gas accretion. We also consider lower solid planetesimal surface densities (4~g\,cm$^{-2}$ at 5.2~AU and 6~g\,cm$^{-2}$ at 4~AU) than most previous calculations (10~g\,cm$^{-2}$ at 5.2~AU) to attain lower heavy element core masses. Until recently, high planetesimal surface densities (about 3 times the minimum mass solar nebula at 5.2 AU) were needed to model Jupiter formation on a reasonable time scale. Advances in the modeling of grain physics \citep{MovshovitzEt2010Icarus} now allow for a reasonable formation time for Jupiter, even with $\sigma= 4~{\rm g\,cm^{-2}}$ considered here. 

We supplement the detailed core nucleated accretion calculations with equilibrium models of Neptune-size planets having H/He envelopes calculated following the approach of \citet{Rogers&Seager2010ApJ} and \citet{Rogers&Seager2010bApJ}. The equilibrium model is less computationally time consuming and allows us to more comprehensively sample the parameter space (heavy element core masses, envelope masses, irradiation levels, and intrinsic planet luminosities) of interest. We focus on low-density planets having equilibrium temperatures of 500--1000~K, since these temperatures are relevant to the planet candidates found in {\it Kepler's} first quarter data \citep{BoruckiEt2011ApJ}. 

We also explore outgassing during planet formation as a possible origin pathway for low-density Neptune-size planets, and derive a limiting mass-radius ($M_p-R_p$) relation bounding the maximum radius/minimum density for planets with primary de-gassed envelopes. Following \citet{ElkinsTanton&Seager2008aApJ}, we consider outgassing of hydrogen gas produced when water reacts with metallic Fe in accreting materials during planet formation. Our outgassed exoplanet models self-consistently treat the connection between the planets' interior structure (iron core mass and mantle composition) and the mass of H$_2$ degassed. We thereby provide the first exoplanet $M_p-R_p$ relations that include the effect of an outgassed H$_2$ gas layer. To derive the limiting $M_p-R_p$ relation, we study planets that accreted from a mixture of water and material with chemical composition characteristic of the high-iron enstatite (EH) chondrite meteorite class, corresponding to end member scenarios yielding maximum outgassed H$_2$.

We begin by describing our equilibrium model for low-mass planets with gas layers in Section~\ref{sec:eqModel}. This model is applied in later sections to explore the mass-radius ($M_p-R_p$) relationships for low-mass planets with voluminous atmospheres of light gases acquired by core nucleated accretion and outgassing of hydrogen. Section~\ref{sec:pform} describes the formation and properties of Neptune-size planets that assembled through core accretion, and Section~\ref{sec:outgas} describes the formation and properties of planets that outgassed hydrogen from dissociated ices. We consider mass loss from the envelope in Section~\ref{sec:mdot}.  We discuss our results and conclusions in Section~\ref{sec:discuss}. 

\section{Models of Planets in Equilibrium: Methods}
\label{sec:eqModel}

We use equilibrium models---spherically symmetric planets in hydrostatic equilibrium---for two applications. The first (Section~\ref{sec:pform}) is to explore the mass-radius relationships for low-mass planets formed via core nucleated accretion. The second application is to again study mass-radius relationships for planets that acquired an envelope of light gases through outgassing (Section~\ref{sec:outgas}).

Our equilibrium model is based upon the planet interior model from \citet{Rogers&Seager2010ApJ, Rogers&Seager2010bApJ}. We have, however, included updates to the temperature profile in the radiative regime of the envelope, and to the outer boundary conditions of the planet. We use equilibrium models to study instantaneous states of evolving planets assuming the planets are undergoing quasistatic evolution. Our work does not focus on cases where the envelope dynamics or variations in the interior luminosity profile have an important effect. 

We assume spherically symmetric and differentiated planets consisting of up to four layers. From the inside out, these layers are an iron core, a silicate mixture, H$_2$O, and a gas envelope. The coupled differential equations describing the mass of a spherical shell in hydrostatic equilibrium:
\begin{eqnarray}
\frac{dr}{dm}&=&\frac{1}{4\pi r^2\rho}\label{eqn:dr}\\
\frac{dP}{dm}&=&-\frac{Gm}{4\pi r^4}\label{eqn:dP},\label{eq:dPdr}
\end{eqnarray}
\noindent and the differential equation describing the radial optical depth, $\tau$, in the gas layer
\begin{equation}
\frac{d\tau}{dm}=-\frac{\kappa}{4\pi r^2}.
\label{eqn:dtau}
\end{equation}
\noindent are integrated inward from the top of the planet's envelope. Above, $m$ is the interior mass coordinate, $r$ is the distance from the planet center, $P$ is the pressure, $\rho$ is the mass density, $\kappa$ is the mean opacity at thermal wavelengths, and $G$ is the gravitational constant.

Within each chemical layer, the equation of state (EOS) relates the density $\rho\left(m\right)$ to the pressure $P\left(m\right)$ and temperature $T\left(m\right)$. In analogy to the models in Section~\ref{sec:pform}, we define the exterior boundary condition on the planet envelope ($r=R_p$, $m=M_p$, $\tau=\tau_R$, $P=P_R$) at radial optical depth $\tau_R=2/3$. We then determine the corresponding pressure $P_R$ by imposing
\begin{equation}
P_R =\frac{g\tau_R}{\kappa_R},
\label{eq:sbcP}
\end{equation}
\noindent where $\kappa_R$ is the Rosseland mean opacity at the photosphere boundary, and $g=GM_p/R_p^2$ is the gravitational acceleration. While the density $\rho$ varies abruptly between the chemical layers, both $P$ and $T$ are continuous across layer boundaries.  For a specified planet composition, energy budget, and mass, $M_p$, the planet radius, $R_p$, is iterated until a self-consistent solution satisfying the inner boundary condition ($r=0$, $m=0$)  is achieved. 

We assume that the gas envelope is in radiative-convective equilibrium, with an outer radiative zone surrounding a convective layer at greater depths. Within the thin outer edge of the envelope, we adopt the isotropic average temperature profile from Equation (29) in \citet{Guillot2010A&A},
\begin{eqnarray}
T\left(\tau\right) &=&\frac{3T^4_{int}}{4}\left[\frac{2}{3}+\tau\right]\label{eq:Trad}\\
&&+\frac{3T^4_{irr}}{4}f\left[\frac{2}{3}+\frac{1}{\gamma\sqrt{3}}+\left(\frac{\gamma}{\sqrt{3}}-\frac{1}{\gamma\sqrt{3}}\right)e^{-\gamma\tau\sqrt{3}}\right],\nonumber
\end{eqnarray}
%\begin{equation}
%T\left(\tau\right)=\frac{3T^4_{int}}{4}\left[\frac{2}{3}+\tau\right]\label{eq:Trad}+\frac{3T^4_{irr}}{4}f\left[\frac{2}{3}+\frac{1}{\gamma\sqrt{3}}+\left(\frac{\gamma}{\sqrt{3}}-\frac{1}{\gamma\sqrt{3}}\right)e^{-\gamma\tau\sqrt{3}}\right],
%\end{equation}
\noindent an analytic solution to the ``two-stream"  gray equations of radiative transfer for a plane-parallel irradiated atmosphere. The irradiation temperature, $T_{irr}$, characterizes the short wave energy flux received by the planet from the star  and relates through the redistribution factor, $f$, to the equilibrium temperature of the planet in the radiation field of the star $T_{eq}=f^{1/4}T_{irr}$ ($f=1/4$ for full redistribution). The intrinsic temperature $T_{int}=\left(L_p/4\pi R_p^2\sigma_B\right)^{1/4}$ parameterizes the total intrinsic luminosity of the planet, $L_p$ ($\sigma_B$ denotes the Stefan-Boltzmann constant). The total intrinsic planet luminosity, $L_p$, is the sum total of contributions from envelope contraction, radioactive decay, and secular cooling of the core. The ratio of the short-wave and long-wave optical depths is  represented by $\gamma$. We take $\gamma=0.6\sqrt{T_{irr}/2000~{\rm K}}$, which \citet{Guillot2010A&A} found provided a good match to detailed calculations of hot Jupiter atmospheres from \citet{FortneyEt2008ApJ}. 

In highly irradiated planet atmospheres, the radiative regime of the envelope may extend to depths beyond where the plane parallel approximation (assumed when deriving Equation~\ref{eq:Trad}) is valid. In these cases, once all of the incoming stellar radiation is absorbed at optical depths $\tau\gg1/\sqrt{3}\gamma$, we transition smoothly to the radiative diffusion equation,
\begin{equation}
\frac{dT}{dr}=-\frac{3\kappa\rho}{16\sigma_BT^3}\frac{L_p}{4\pi r^2}.
\label{eq:dTraddiff}
\end{equation}
\noindent The onset of convective instabilities $\left(0<\left({\partial \rho}/{\partial s}\right)_P{ds}/{dm}\right)$ determines the depth of the transition to the convective layer of the gas envelope. In the convective regime, we adopt an adiabatic temperature profile. 

We use the EOS from \citet{SaumonEt1995ApJS}. The effect of uncertainties in the H/He EOS \citep[see e.g.,][]{MilitzerEt2008ApJ, NettelmannEt2008ApJ}  will be small compared to the effect of uncertainties in the heavy element composition and distribution, for the low-mass planets we are considering. While the major uncertainties in the EOS are at Mbar pressures or above, the pressures at the base of our H/He envelopes are typically less than a few tenths of a Mbar.
As in the formation and evolution models of Section~\ref{sec:pform}, we use Rosseland mean molecular opacities from \citet{FreedmanEt2008ApJS}. We neglect grain opacities in our equilibrium models, however, since we are interested in the planet radii at late times, after all the grains have settled. 

Under the H/He envelope, the rock-ice interior is modeled with differentiated layers of iron, Fe$_{0.1}$Mg$_{0.9}$SiO$_3$ silicates, and H$_2$O.  For these materials, we employ EOS data sets from \citet{SeagerEt2007ApJ}, which were derived by combining experimental data at $P\lesssim200~\rm{GPa}$ with the theoretical Thomas-Fermi-Dirac (TFD) equation of state at high pressures, $P\gtrsim10^4~\rm{GPa}$. The equation of state at intermediate pressures between $\sim200$ and 10,000~GPa is not well known, since this pressure range is neither easily accessible by experiments nor by TFD theory. For H$_2$O, \citet{SeagerEt2007ApJ} used density functional theory calculations to fill in the EOS in this pressure regime, while for all other materials, they bridged the pressure gap by extrapolating the empirical Birch-Murnagham EOS and the theoretical TFD EOS to higher and lower pressures, respectively. Thermal effects are neglected in the \citet{SeagerEt2007ApJ} EOSs -- at the high pressures found in the interior layers, thermal corrections have only a small effect on the density, $\rho$. An improvement over our previous models is that we now more consistently take into account the Si/Mg/Fe ratios in the mantle by calculating EOSs for mixtures of MgO, FeO and SiO$_2$. Core mass-radius relations calculated following this scheme (but neglecting the small contribution to pressure from the gaseous envelope) were also employed in the planet evolution calculations of Section~\ref{sec:pform}.

A major uncertainty in the validity of the models comes from the assumption that the layers of water and H (or H/He) are not mixed. For the $T_{eq}=500-1000$~K planets considered in this work, H$_2$ and H$_2$O are miscible at the pressures and temperatures in the model envelopes. So they could, in principle, be homogeneously mixed. By considering the extreme where the H and H$_2$O are fully separated, we set an upper bound on the planet radius; typically if hydrogen is mixed into the water layer one expects the planet's radius to be smaller \citep[e.g.,][]{NettelmannEt2011ApJ}. Although our aim is to model H/He envelope planets, some of our models do have significant water content, and future work should include the miscibility of H$_2$ and H$_2$O. 

The planet radii in both our equilibrium and evolution models underestimate the planet radii measured during transit in a predictable way. This ``transit radius effect" \citep{BaraffeEt2003A&A, BurrowsEt2003ApJ} is a consequence of our exterior boundary condition (Equation~\ref{eq:sbcP}), which pegs our model planet radii, $R_p$ at a radial optical depth $\tau_R=2/3$. In contrast, it is the transverse optical depth for transmission through the planet limb that determines the transit radius. \citet{Hansen2008ApJS} derived a correction for the transit radius effect,
\begin{equation}
\Delta R = H_R\ln\left[\gamma\tau_R\left(\frac{2\pi R_T}{H_R}\right)^{1/2}\right],
\label{eq:tre}
\end{equation}
\noindent where the transit radius $R_T=R_p+\Delta R$ is defined at a transverse optical depth of unity, and $H_R$ represents the atmospheric scale height at the planet limb. Equation~(\ref{eq:tre}) applies when $H_R\ll R_p$, and assumes the outer limb of the planet atmosphere is well described by an ideal gas. For the low-mass ($M_p<30~M_{\oplus}$) planets with hydrogen-rich envelopes we are considering, the transit radius correction is typically between 1\% and 10\%. Equation~\ref{eq:tre} assumes a clear cloud-free planet atmosphere; high level clouds or hazes could further enhance the transit radius effect.

\section[]{Planet Formation by Core-Nucleated Accretion}
\label{sec:pform}

\subsection[]{Methods}
\label{sec:pformmod}

Models for the formation and evolution of a planet consisting of a  core of 
heavy elements
and a gaseous envelope of solar composition are calculated according to the
procedures described by \citet{PollackEt1996Icarus} and \citet{MovshovitzEt2010Icarus}.
 The formation calculation consists of
three major parts: ($\it i$) the accretion rate of planetesimals onto the planet; ($\it ii$)
the interaction of the infalling planetesimals with the gaseous envelope; and ($\it iii$)
 the evolution of the gaseous envelope and the determination of the
gas accretion rate. 

The planetesimal accretion rate onto the planetary embryo is based on the equation
 originated by \citet{Safronov1969book}. 
If $M_{\rm core}$ is the mass of the
embryo, then the fundamental equation for its growth is
\be
\frac{dM_{\rm core}}{dt} =  \pi R_{\rm capt}^2 \sigma\Omega F_g , 
\label{eq:saf}
\ee
where $\pi R_{\rm capt}^2$  is the effective geometrical capture 
cross-section, $\sigma$ is the surface density of solid material (planetesimals), 
 $\Omega$ 
is the orbital frequency, and the value of  the
gravitational enhancement factor, $F_g$, is obtained from 
\citet{Greenzweig&Lissauer1992Icarus}, assuming a planetesimal radius of
100 km.  If no gaseous envelope is present, then
$ R_{\rm capt} =  R_{\rm core}$, the heavy element core radius. 
As in our previous publications,  we take the feeding zone from which the
embryo can accrete planetesimals to extend 4 Hill sphere radii
on either side of the orbit \citep{Kary&Lissauer1994nsa}, and assume that the solid surface density $\sigma$
is constant within that zone. The value of                    
$\sigma$ in the feeding zone is adjusted at each time step to 
take into account depletion of planetesimals
by accretion onto the embryo and expansion of the feeding zone  into
undepleted regions,  as the embryo's mass increases. 

The second element of the code calculates the  interactions
between planetesimals and the gaseous envelope of the protoplanet
as they fall through it \citep{PodolakEt1988Icarus}. The details of how
this calculation is performed are described in \citet{PollackEt1996Icarus}, 
\citet{HubickyjEt2005Icarus}, and \citet{MovshovitzEt2010Icarus}.  These     
calculations provide the effective capture radius $R_{\rm capt}$ to be 
used in Eq. (\ref{eq:saf}), as well as the deposition of mass and energy
as a function of radius in the gaseous envelope.  
The effective capture radius can be several times larger than the actual solid heavy element core
radius because of the effects of the gas on slowing down, ablating, and 
fragmenting  the
planetesimals.  It is assumed that the material from the planetesimals
that is deposited in the envelope later sinks to the heavy element core,
releasing gravitational energy in the process \citep{PollackEt1996Icarus}.
This assumption is not entirely accurate: 
\citet{Iaroslavitz&Podolak2007Icarus} show that some of 
the heavy-element material should actually dissolve in the envelope. 
Thus the `core masses' that we calculate actually represent the total
heavy element abundance in the planet in excess of the solar metal
abundance; most of these heavy elements (including all of the rock and organic compounds) would be expected to reside in the actual core of the planet.

The third element of the simulation
is the solution of the four differential equations
of stellar structure for the gaseous envelope, with energy sources from
accreting planetesimals, from gravitational contraction, and from cooling.
The adiabatic
temperature gradient is assumed in convection zones. 
 At the heavy element core boundary, the luminosity $L_r$ is set to the energy
deposition rate for the planetesimals that hit the heavy element core.
Outside the heavy element core, the energy supplied by ablated and fragmented
planetesimals is included as a source term in the energy equation.

At the core-envelope
interface, the radius is set to that of the outer edge of the heavy element core.
The heavy element core mass-radius relation is calculated using the equilibrium model described in Section~\ref{sec:eqModel}, assuming 10\% Fe, 23\%  Fe$_{0.1}$Mg$_{0.9}$SiO$_3$, and 67\% H$_2$O, by mass. The heavy element core composition we adopt is motivated by comet compositions, and represents 
rock with a Fe/Si ratio near solar mixed with ice in a ratio of 1:2 by mass. 

At the surface, gaseous mass of solar composition is added at a sufficient rate to maintain
an outer radius $R_p = R_{\rm eff}$, which is given by \citep{BodenheimerEt2000Icarus} as
\begin{equation}
 R_{\rm eff} = \frac{G M_p}{c_s^2 + \frac{G M_p}{K R_H}},
\label{eq:bh}
\end{equation}
where $R_H$ is the Hill sphere radius, $c_s$ is the sound speed in the disk
outside the planet,  $M_p$ is the total planet mass, and, nominally, 
$K=1$.  Note that when
$R_H$ is large compared with the Bondi accretion radius,  $GM_p/c_s^2$,  the
expression reduces to the Bondi radius, while in the case of the opposite
limit, $R_{\rm eff} \rightarrow R_H$.  
 In developments after the above expression was formulated, it turned out
that $K$ had to be modified. 
Three-dimensional calculations of disk-planet interaction \citep{LissauerEt2009Icarus} gave the result that not all the gas passing through the Hill
sphere is actually accreted by the planet; some of it simply flows through
and rejoins the disk's azimuthal motion. The 3D simulations provided an
estimate of the effective planetary radius, which corresponds to $K=0.25$, the
value used in this paper.  

The density and temperature at the planet's surface are set to
assumed nebular values $\rho_{\rm neb}$, $T_{\rm neb}$, respectively.
The value of  $T_{\rm neb}$ is  constant in time, while $\rho_{\rm neb}$
decreases linearly to zero  with time, over a time scale $T_d \approx 2-3$ Myr.
In a variation of this boundary condition,  $\rho_{\rm neb}$ is constant
in time up to a time comparable to $T_d$, then it is linearly reduced to zero
on a time scale of $10^5$ yr.  These
assumptions roughly characterize the dissipation of the gaseous disk. $T_{neb}$ is held constant while the planet is accreting; our model incorporates migration only through temperature increases subsequent to the conclusion of the planet's growth.  Modeling simultaneous migration and accretion is beyond the scope of this work.

When $\rho_{\rm neb}$ approaches zero, the accretion of gas is halted
and the evolution is calculated at constant mass over time scales up to
3--4 Gyr. The envelope mass at cutoff in these simulations is always
small enough that rapid runaway gas accretion does not occur, and
Eq. (\ref{eq:bh}) is always valid for the determination of the gas
accretion rate. The accretion rate required to keep $R_p = R_{\rm eff}$
 remains much lower than the
limit imposed by  disk physics  in supplying material to the Hill
sphere of the planet \citep{LissauerEt2009Icarus}. Once gas accretion is shut
off, $R_p$ rapidly falls below $R_{\rm eff}$, and the planet becomes
isolated from the disk. The surface boundary condition changes to that
of a hydrostatic atmosphere that radiates from the photosphere:
\begin{subequations}
\label{eq:sbc}
\begin{equation}
L = 4 \pi R_p^2 \sigma_B T_{\rm eff}^4 ~~~{\rm and}
\end{equation}
\begin{equation}
\kappa P = \frac{2}{3} g,
\end{equation}
\end{subequations}
where $\sigma_B$ is the Stefan-Boltzmann constant, $T_{\rm eff}$ is
the surface temperature, $L$ is the total luminosity (energy 
radiated per second)
of the planet, and $\kappa$, $P$, and $g$ are, respectively, the
Rosseland mean opacity, the pressure, and the acceleration of 
gravity at the photosphere.  There are two contributions to 
$T_{\rm eff}$: that from the internal luminosity provided by the
planet, and that from the energy absorbed from the central star
and re-radiated by the planet. Thus:
\begin{equation}
T_{\rm eff}^4 =  T_{\rm int}^4 + T_{\rm eq}^4,
\end{equation}
where $T_{\rm int}$ is the internal contribution (generally small), and
$ T_{\rm eq}$ is the equilibrium temperature of the planet in the
radiation field of the star. The former quantity is determined from the
evolutionary calculation, while the latter is a parameter that depends
on the assumed distance of the planet from the star and the stellar
luminosity.

The equation of state of the gas is taken from \citet{SaumonEt1995ApJS},
interpolated to our assumed composition of hydrogen mass fraction
X~=~0.74, helium mass fraction Y~=~0.243, and metal mass fraction 
${\rm Z}=1 - {\rm X} - {\rm Y} = 0.017$. Although the equation of state in the outer, low-density
layers is essentially that of an ideal gas, the inner regions near the heavy element core can be significantly non-ideal once the envelope has become   
sufficiently compressed.

The Rosseland mean opacity calculation has three 
components. At temperatures above 3000 K, the molecular/atomic  opacities of
\citet{Alexander&Ferguson1994ApJ} are used. In practice the details of the
opacities in this region are unimportant because the energy
transport is almost always by convection. In the temperature range
100--3000 K the molecular opacities, without grains, of \citet{FreedmanEt2008ApJS} are used. Grain opacities are then added
in the temperature range 100--1800 K.  Two sources of grains are taken
into account, first, those provided by the ablating planetesimals as 
they interact with the envelope, and, second, those that accrete along with
the gas at the surface of the planet. At each time step of the
evolutionary calculation, and at each depth in the envelope,
the grain size distribution is recalculated, taking into account the
coagulation and settling of grains.  The size distribution is
represented by 34 bins, covering the size range 1.26~$\mu$m to 
2.58~mm. The effective cross-sections for
absorption and scattering are calculated  as a function of grain size
and frequency; then an integration over grain size and frequency gives
the Rosseland mean opacity as a function of depth. The details of the
grain physics are given in \citet{Movshovitz&Podolak2008Icarus} and 
\citet{MovshovitzEt2010Icarus}.  The grains are composed purely of silicates,
with a dust-to-gas ratio of about 0.01 by mass; 
little error results from this assumption compared to the
uncertainties in grain shape, sticking probability, and radiative properties \citep{MovshovitzEt2010Icarus}. 
Grains are assumed to be completely evaporated above 1800 K.  The grains are 
important during the gas accretion phase. Once accretion is shut off,
the grains rapidly settle towards the center and are evaporated. This effect is included in the calculations and indicates that any grains remaining in the atmosphere have a negligible effect upon the evolution. Thus in the final constant-mass evolution phase, the molecular opacities
completely dominate.

\begin{figure}[htb]
\begin{center}
%\epsscale{0.8}
\plotone{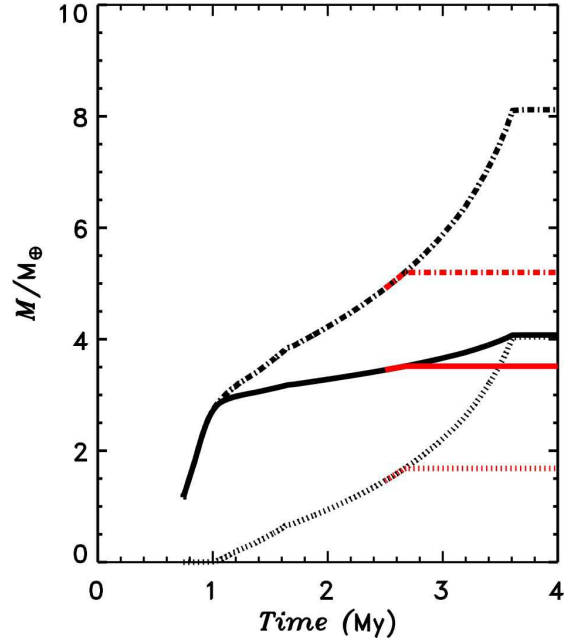}
%\plotone{run4bgfnt3.eps}
\caption{Mass of the protoplanet as a function of time for Runs   I.
 For Run Ia (black curves) the solid line denotes the
mass of the heavy element core, the dotted line the mass of the H/He envelope, and
the short-dash-dot line the total mass.  For Run Ib, the same line types
are shown in red.
}
\label{fig:f1}
\end{center}
\end{figure}

\begin{figure}[htb]
\begin{center}
%\epsscale{0.8}
\plotone{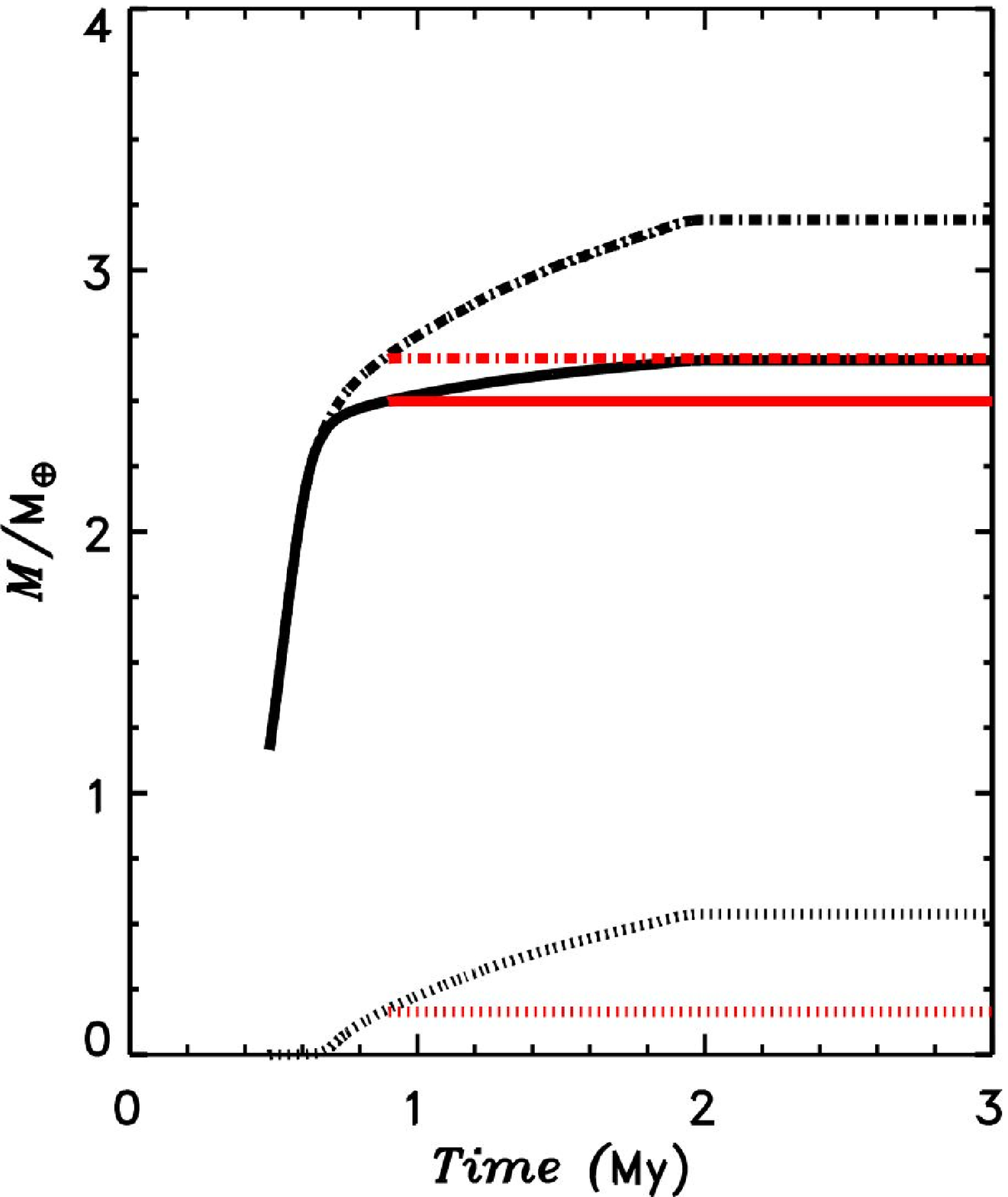}
%\plotone{run5bgfnt3.eps}
\caption{Mass of the protoplanet as a function of time for Runs  II.
 For Run IIa (black curves)  the solid line denotes the
mass of the heavy element core, the dotted line the mass of the H/He envelope, and
the short-dash-dot line the total mass.  For Run  IIb, the same line types
are shown in red.
}
\label{fig:f2}
\end{center}
\end{figure}

%%%%%%%%%%%%%%%%%%%%%%%%%%%%%%%%%%%%%%%%%%%%%%%%%%%%%%%%%%%%%%%%%%%
\begin{table*}
 \caption{Input Parameters for Evolutionary Runs}\label{table:1}
 \centering
 \vspace*{1ex}%
% \resizebox{1.0\textwidth}{!}{%
 \small
 \begin{tabular}{|l||cccccc|}
 \hline
 Run    
 & $a$ (AU)
 & $\sigma$ (g/cm$^2$)                & $\sigma_{XY}$  (g/cm$^2$)  
 & $T_{\mathrm{neb}}$ (K)  &  $T_d$ (Myr) 
 & ${M}_{\rm iso}$ ($M_\oplus$)   
   \\
 \hline\hline
 Ia   
 & 5.2
 &          4   & 280                       
 &  115                    & 3.5        
 &  $2.92$    
  \\
 \hline
 Ib   
 & 5.2
 &          4                  & 280    
 &  115                    &  2.5
 &  $2.92$                 
    \\
 \hline
 IIa      
 & 4.0
 &     6                      & 420    
 &  125                    &  2.0         
 &  $2.42 $     
   \\
 \hline
 IIb         
 & 4.0 
 &     6   & 420    
 &  125                    & 0.9       
 &  $2.42$ \\                 
 \hline

 \end{tabular}
    %   }
\end{table*}
%%%%%%%%%%%%%%%%%%%%%%%%%%%%%%%%%%%%%%%%%%%%%%%%%%%%%%%%%%%%%%%%%%%

%%%%%%%%%%%%%%%%%%%%%%%%%%%%%%%%%%%%%%%%%%%%%%%%%%%%%%%%%%%%%%%%%%%
\begin{table*}
 \caption{Results from Evolutionary Runs: Masses and Radii}\label{table:2}
 \centering
 \vspace*{1ex}%
% \resizebox{1.0\textwidth}{!}{%
 \small
 \begin{tabular}{|l||ccccccc|}
 \hline
 Run    
 & $M_p$ ($M_\oplus$)
 & $M_{\rm core}$ ($M_\oplus$)
 & $M_{\rm env}$ ($M_\oplus$)            & $R_{1;500}$ ($R_\oplus$)  
 &  $R_{4;500}$ ($R_\oplus$)  &  $R_{1;1000}$ ($R_\oplus$) 
 & $R_{4;1000}$ ($R_\oplus$)   
   \\
 \hline\hline
 Ia   
 & 8.13
 & 4.08
 &        4.05  &  9.8                      
 &  8.1                    & 14.8       
 &  $11.6$    
  \\
 \hline
 Ib   
 & 5.20
 & 3.52
 &         1.68                & 8.0    
 &  6.6                    &  15.7
 &  $11.5$         
    \\
 \hline
 IIa      
 & 3.19 
 & 2.65 
 &     0.54                   & 6.0   
 &  5.0               &  17.9         
 &  11.7        
   \\
 \hline
 IIb          
 & 2.66 
 & 2.50 
 &     0.16   & 3.6    
 &  3.3                    & 6.7       
 &  6.2    \\                 
 \hline

 \end{tabular}
    %   }
%%%%%%%%%%%%%%%%%%%%%%%%%%%%%%%%%%%%%%%%%%%%%%%%%%%%%%%%%%%%%%%%%%%
 \hspace{0.2in}%
 \begin{minipage}[t]{\linewidth}
 \footnotesize
 \noindent%
 \vskip 0.12in
 The first subscript on the radius gives the evolutionary time in 
  Gyr. The second subscript gives the assumed equilibrium temperature of the
  planet.
 \end{minipage}
 \label{tab:Rp}
\end{table*}
%%%%%%%%%%%%%%%%%%%%%%%%%%%%%%%%%%%%%%%%%%%%%%%%%%%%%%%%%%%%%%%%%%%

\subsection[]{Evolution Input Parameters and Results}
\label{sec:pformres}

The planet initially consists of a heavy element core of 1 $M_\oplus$ and a light element
envelope of about $10^{-5}$ $M_\oplus$. 
The protoplanet is located at either 5.2 AU or 4.0  AU 
in a protoplanetary disk, with
the solid surface density $\sigma = 4$ g cm$^{-2}$ at 5.2 AU and 
6   g cm$^{-2}$ at 4 AU.  
The initial evolutionary time is set to
$7.3 \times 10^5$ yr and $4.8 \times 10^5$ yr, respectively, for $\sigma=4$ 
and 6 g cm$^{-2}$, approximately the time needed to assemble a heavy element core of mass $M_{core}=1~$$M_\oplus$.

The quantity $T_{\rm neb}$ is set to 115 K at 5.2 AU
and 125 K at 4 AU.
 Then $\rho_{\rm neb}=\sigma_{XY}/(2H)$, where $\sigma_{XY}=70 \sigma$ is 
the surface density of the gas component. As
mentioned above,  $\rho_{\rm neb}$ in general declines with time.
The scale height of the gas in the disk  $H=0.05 a$, where $a$ is the orbital distance
from the star. Once started, the evolution consists of three main
phases. The first involves primarily accretion of solids onto the heavy element core, with a
relatively low-mass envelope and a low gas accretion rate. The solids
accretion rate slows down significantly near the point where the isolation
mass  ($M_{\rm iso}$) for the core is reached; for $\sigma=4$ g cm$^{-2}$ at 5.2 AU 
this mass is
about 2.9  $M_\oplus$ and for $\sigma=6$ g cm$^{-2}$ at 4 AU, about 2.4  $M_\oplus$. During the second phase, the gas accretion rate is
about 3 times as high as the core accretion rate, $\dot{M}_{env}\approx 3\dot{M}_{core}$, and both are nearly
constant in time \citep{PollackEt1996Icarus}. The envelope mass builds up relative to the heavy element core mass,
which grows slowly. The phase of  rapid gas accretion, which for giant planets begins 
once the envelope mass $ M_{\rm env}$ becomes about equal to  $M_{\rm core}$,
 does not occur in these calculations.  Instead, gas accretion is cut off
and the planet evolves through a third phase at constant mass with
boundary conditions provided by Eq. (\ref{eq:sbc}). During the early part
of this phase, the planet is assumed to migrate to a position within
1 AU from the star. Representative cases with $T_{\rm eq} = 500$ K and 
1000 K are presented.

The input parameters of the four
runs are shown in Table 1, which includes $\sigma$, the gas
surface density $\sigma_{XY}$,  the surface boundary temperature 
$T_{\rm neb}$, 
and the isolation mass $M_{\rm iso}$.

\begin{figure}[htb]
\begin{center}
%\epsscale{0.8}
\plotone{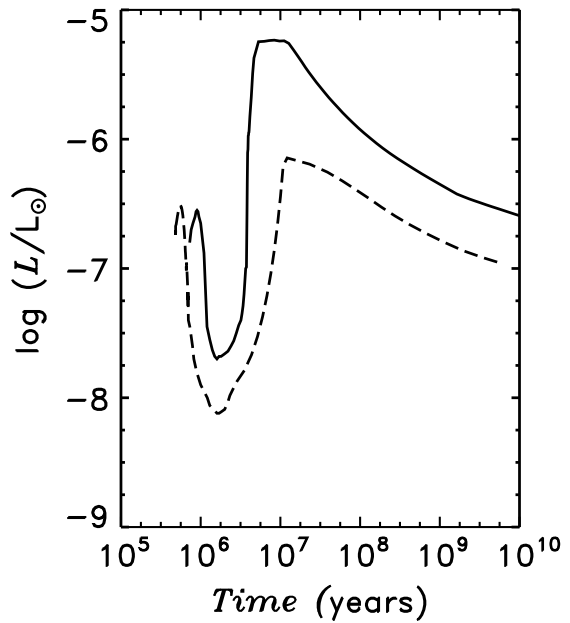}
%\plotone{plum4.eps}
\caption{The protoplanet's total luminosity, including internal and irradiation contributions, as a function of time during the 
 formation phase and the contraction/cooling phase for Run Ia 
 ({\it solid curve}), and  
Run IIa ({\it dashed curve}). The equilibrium temperature is increased
to 500 K, after the formation phase,  during these runs.
}
\label{fig:f3}
\end{center}
\end{figure}

The results of our calculations  for Runs  Ia, Ib, IIa,  and  IIb
 are shown in Figures \ref{fig:f1}, \ref{fig:f2},  and
\ref{fig:f3}. The masses and radii that are derived for the four runs
are listed in Table 2.

Run Ia is based on a disk with  $\sigma=4$ g cm$^{-2}$ at 5.2 AU.
This value is only
slightly greater than that of the minimum mass solar nebula.  But note 
that our calculation of $F_g$ (Eq. \ref{eq:saf}) neglects transport of solids into or away from
 the planet's accretion zone. Moreover, our planetesimals are
all assumed to have the same radius, 100 km. In fact there must be a range of
planetesimal sizes, and the effective planetesimal size is not well known. Smaller planetesimals would result in more rapid accretion \citep[see footnote 3 of][]{LissauerEt2009Icarus}. The accretion rate that is
actually calculated  may thus correspond to a value of $\sigma$ 
slightly different from 4  g cm$^{-2}$.

The details of the calculation with the parameters of Run Ia are
presented in \citet{MovshovitzEt2010Icarus},  their Run $\sigma$4.  In that
paper the run is continued well into the phase of rapid gas accretion,
and is terminated with $M_{\rm core} = 4.74 $ $M_\oplus$ and 
$M_{\rm env} = 34 $ $M_\oplus$. The formation time for a giant planet
is found to be 4 Myr. In the present run, the accretion of gas and solids
is cut off at a time of 3.5 Myr, consistent with estimated lifetimes
of protoplanetary disks \citep{Hillenbrand2008PhysScr}. At that time the value
of $\rho_{\rm neb}$ is assumed to decrease to zero on a time scale of
10$^5$ yr. The calculation is then continued up to Gyr times with 
constant values of $M_{\rm core}= 4.08$  $M_\oplus$ and 
 $M_{\rm env}= 4.05$  $M_\oplus$. At the beginning of this phase the
equilibrium temperature is gradually increased, on a time scale of 
4 Myr, to an assumed final value of 500 K. A gradual
 increase in $T_{\rm eq}$ 
to  1000~K was accomplished in a total time of  $6\times10^7$~yr. 
The final values of $R_p$ for these two
temperatures and for times of 1 and 4 Gyr are given in Table 2; 
they are close to Jupiter's radius R$_J\approx11~R_\oplus$, even though
the planet's mass is only 8.13 $M_\oplus$.

Run Ib also is based on the run $\sigma$4 from \citet{MovshovitzEt2010Icarus}.
In this case the accretion of gas and solids was cut off at 2.5 Myr, at
which point $M_{\rm core}= 3.52$  $M_\oplus$  and $M_{\rm env}= 1.68$
$M_\oplus$. The evolution was again continued into the phase of
cooling and contraction at constant mass, with assumed values of 
$T_{\rm eq}$ of 500 and 1000 K. In the case with $T_{\rm eq} =1000$~K,
the final radii are again comparable to or larger than  R$_J$. In 
the case with $T_{\rm eq} =500$ K the minimum radius is 6.6~$R_\oplus$, only slightly smaller than the corresponding value in Run Ia.

Run IIa is an entirely new calculation, with the planet forming at 4 AU
in a disk with $\sigma=6$ g cm$^{-2}$. During the initial phase of
rapid core accretion, the luminosity reaches a maximum of 
$3.1 \times 10^{-7}$ L$_\odot$ at a time of $6.2 \times 10^5$ yr.
The heavy element core mass is 2.2 $M_\oplus$ and the core accretion rate
$\dot M_{\rm core} = 5 \times 10^{-6}$ $M_\oplus$ yr$^{-1}$ at this time.
Later, at 1 Myr, $\dot M_{\rm core}$ has decreased to $ 2 \times 
10^{-7}$ $M_\oplus$ yr$^{-1}$, and $\dot M_{\rm env}$ has increased to 
 $ 5 \times 10^{-7}$ $M_\oplus$ yr$^{-1}$. The luminosity has
decreased to $10^{-8}$  L$_\odot$. Because of computational time
limitations, and to obtain a lower envelope mass than that found for
Run Ib, the accretion in this run was cut off at 2 Myr, 
with $M_{\rm core} = 2.65$  $M_\oplus$ and  $M_{\rm env} = 0.54$  $M_\oplus$.
If the evolution had been continued up to 2.5 Myr, the heavy element core mass would
have been practically unchanged, and the $M_{\rm env}$ would have
increased  by about 0.25  $M_\oplus$.
At the end of the contraction/cooling phase the radii are in the
range 5--6  $R_\oplus$ for the case of $T_{\rm eq}=500$~K, and for 
 $T_{\rm eq}=1000$~K they are larger then $R_J$, close to the values
obtained in Runs I for that temperature.

To investigate the effect of an even smaller value of $M_{\rm env}$,
Run IIb was calculated with the same parameters as Run IIa, but with an
arbitrary accretion cutoff at $9.1 \times 10^5$ yr. At that point,
 $M_{\rm core} = 2.5$  $M_\oplus$ and  $M_{\rm env} = 0.16$  $M_\oplus$.
Final radii turned out to be in the range 3--4  $R_\oplus$ for 
$T_{\rm eq} = 500$ K and in the range 6--7  $R_\oplus$ for
$T_{\rm eq} = 1000$ K. The significant reduction in envelope mass resulted in final radii that are about half the values obtained
for Run IIa.

We neglect heating from radioactive decay in the core nucleated accretion calculations. Including this additional energy source would delay envelope contraction and planet cooling. Consequently, the planet radii at 1 Gyr and 4 Gyr in Table 2 may be systematically underestimated by a small amount. We estimate that, for the cases in Table 2, the planet luminosity from radioactive decay would be roughly one order of magnitude smaller than luminosity from envelope contraction, assuming bulk Earth abundances of K, U, and Th in the heavy element cores \citep{VanSchmus1995geph}. The fractional contribution to the planet energy budget from radioactive heating will be higher for older planets (4 Gyr) and cases where the heavy element core contributes a larger fraction of the planet mass (Run II).

\subsection{Equilibrium Model Results}
\label{sec:eqres}

 In this section we explore planet radii over a wide range of heavy element core masses, envelope masses, irradiation levels, and intrinsic planet luminosities. The planet formation and evolution model described in the Section~\ref{sec:pformmod} is computationally intensive. Since it is not feasible to simulate planets under all conditions of interest following that approach alone, we enlist an equilibrium planet structure model (Section~\ref{sec:eqModel}) to cover a wider range of parameter space.

Our equilibrium model shows good agreement with the planet evolution models in Section~\ref{sec:pformres} despite the differences in their treatment of the outer radiative regime, the intrinsic planet luminosity, and the effects of stellar insolation. For each entry in Table~\ref{tab:Rp}, we applied the equilibrium model to simulate the same combination of $M_{core}$, $M_{env}$, $T_{eq}$ and $T_{int}$. The radii at $T_{eq}=500$~K in the two models agree to better than $0.2~R_{\oplus}$ in every case.  The planet radii at $T_{eq}=1000$~K are more sensitive to model assumptions and exhibit larger discrepancies (up to 14\%, with the equilibrium model radii systematically below those in Table~\ref{tab:Rp}). 

\begin{figure*}[htb]
\epsscale{1}
\begin{center}
\plottwo{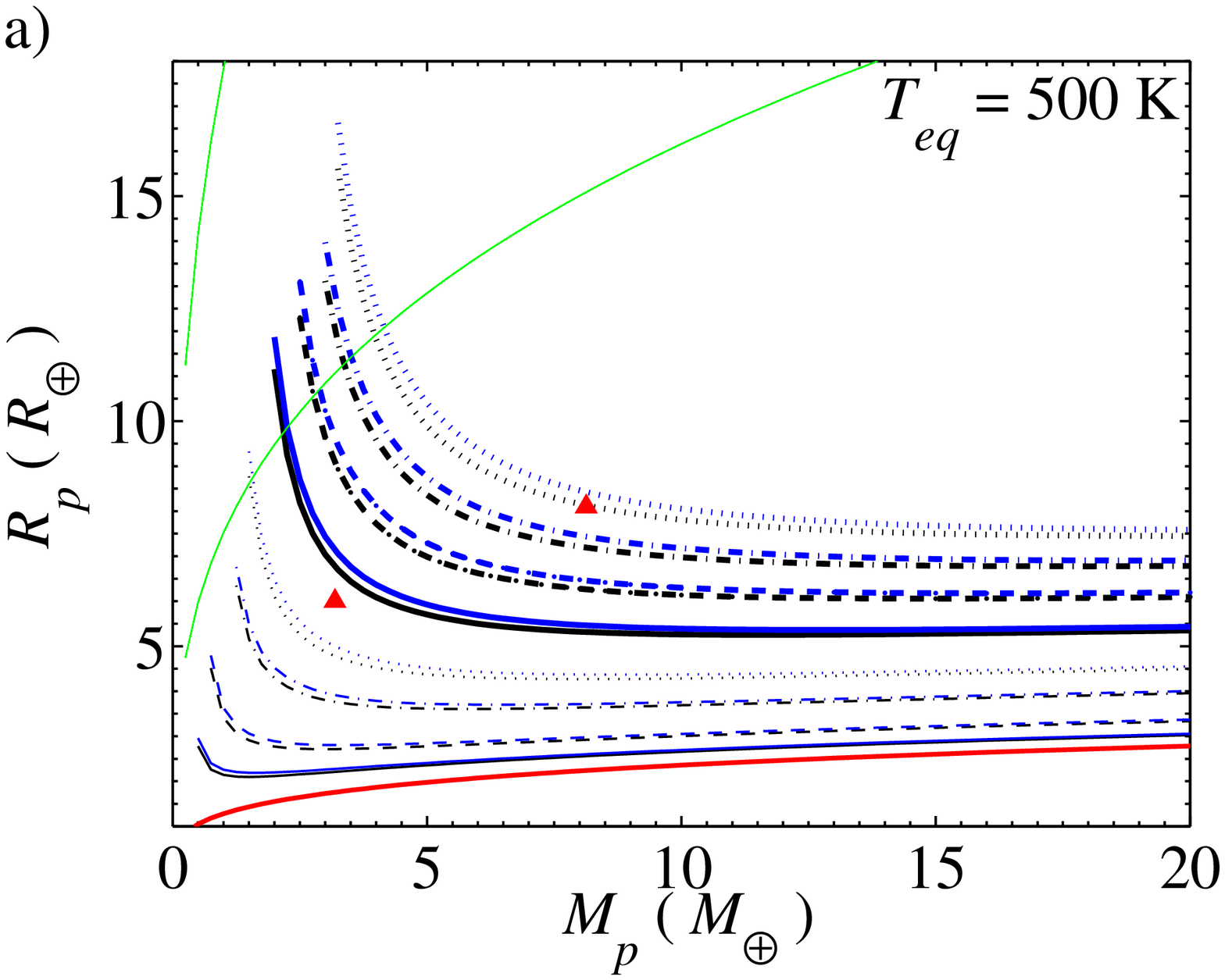}{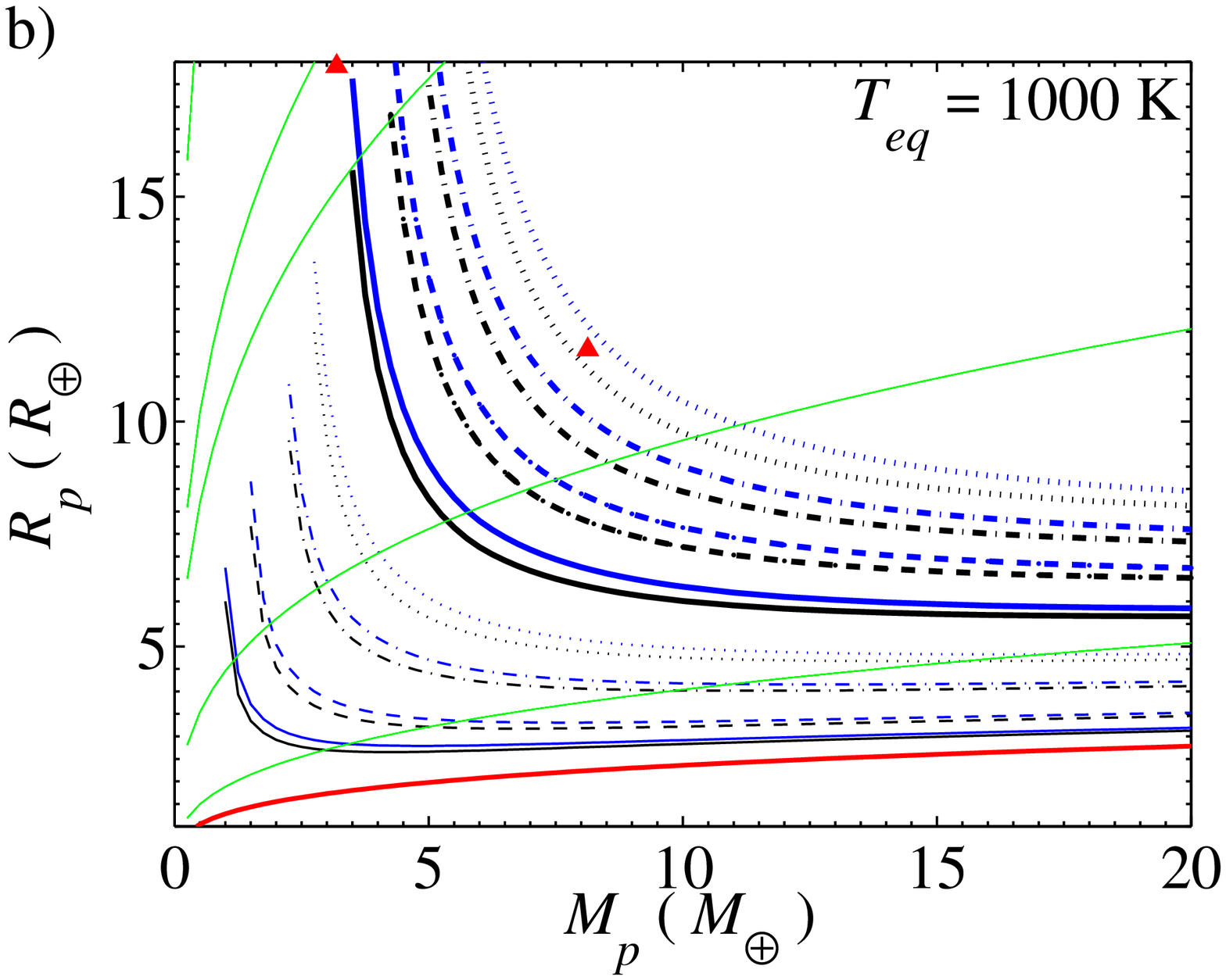}
%\plottwo{lowrho2gGice_wpv36h_Teq500_varygmf.eps}{lowrho2gGice_wpv36h_Teq1000_varygmf.eps}
\caption{Equilibrium mass-radius relations for various choices of envelope mass fraction, $M_{env}/M_p$. All data in this plot have $L_{p}/M_p=10^{-10.5}~\rm{W\,kg^{-1}}$, and (a) $T_{eq}=500$~K or (b) $T_{eq}=1000$~K. Each curve corresponds to a different value of $M_{env}/M_p$: 0.001 (thin solid), 0.01 (thin dashed), 0.05 (thin dot-dashed), 0.1 (thin dotted), 0.2 (thick solid), 0.3 (thick dashed), 0.4 (thick dot-dashed), and 0.5 (thick dotted). Black lines denote our model radii (defined at a radial optical depth $\tau=2/3$), while the corresponding blue lines represent radii corrected for the transit radius effect. The thick red line is the mass-radius relation for icy  heavy element cores having no envelope ($M_{env}=0$). Red triangles present the subset of Table~\ref{table:2} evolutionary run results that have $L_{p}/M_p\approx10^{-10.5}~\rm{W\,kg^{-1}}$: Run Ia ($M_p=8.3~M_{\oplus}$) at 4~Gyr, and Run IIa ($M_p=3.19~M_{\oplus}$) at 1~Gyr. The green curves show the effective planet Roche-lobe radius for four different choices of host-star properties representative of spectral classes M5 V, M0 5V, K0 V, and G2 V  (in order of increasing Roche lobe radii). The K0 V and G2 V Roche-lobe radii are beyond the scale of the $T_{eq}=500$~K plot.}
\label{fig:MRgmf}
\end{center}
\end{figure*}

\begin{figure*}[htb]
\begin{center}
\plottwo{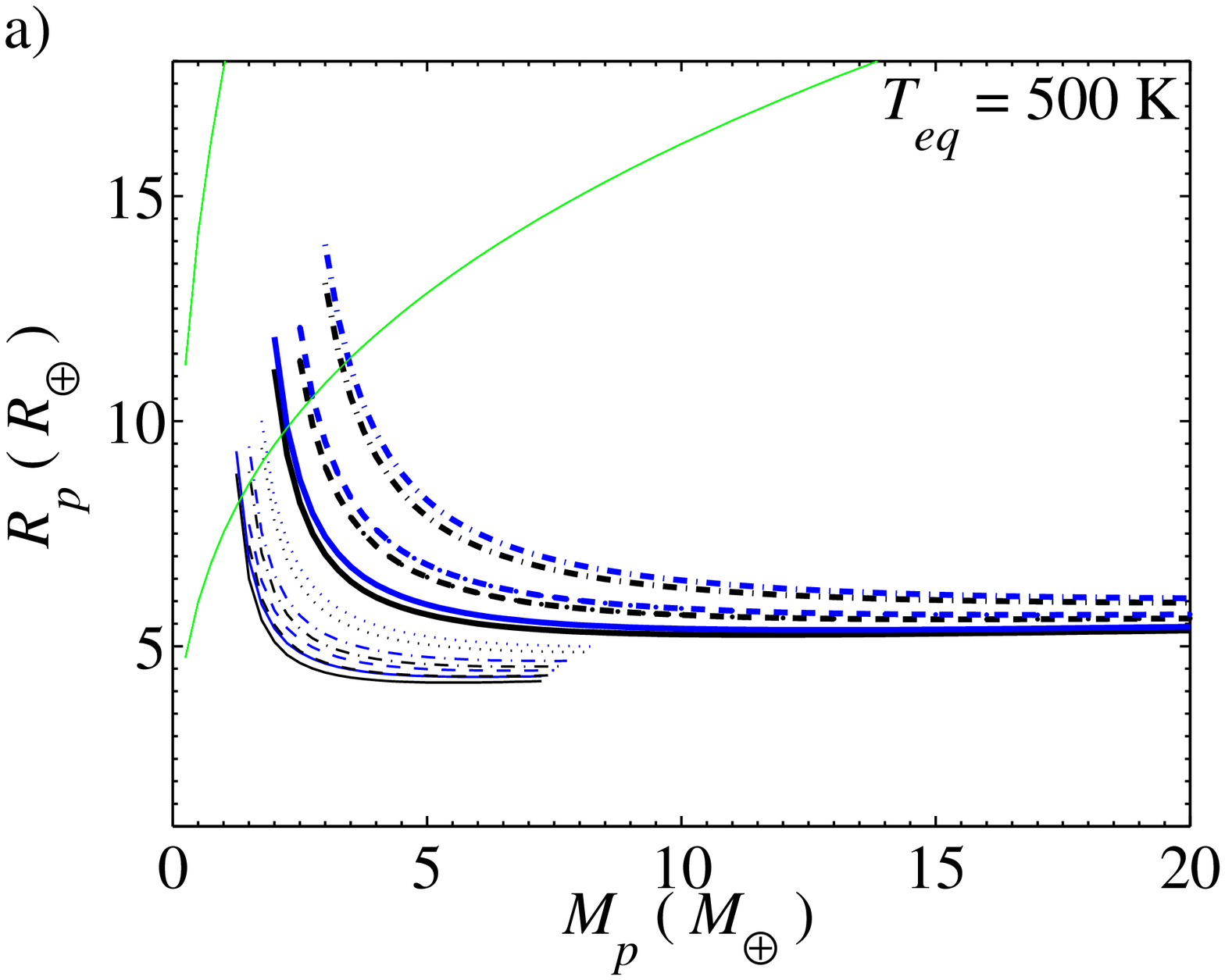}{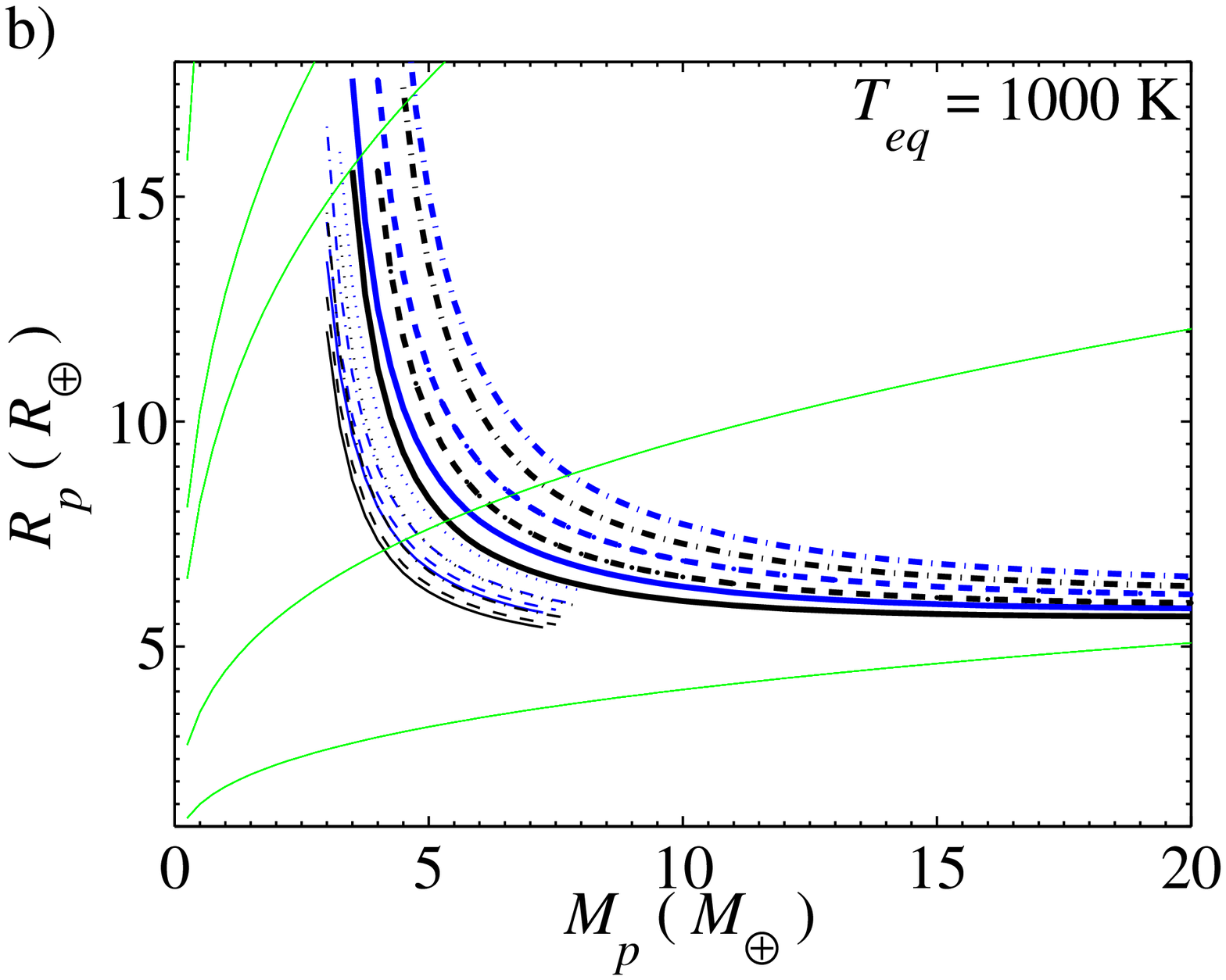}
%\plottwo{lowrho2gGice_wpv36h_Teq500_varyTint.eps}{lowrho2gGice_wpv36h_Teq1000_varyTint.eps}
\caption{Equilibrium mass-radius relations for various choices of intrinsic planet luminosity $L_{p}/M_p$. All data in this plot have $M_{env}/M_p=0.2$, and (a) $T_{eq}=500$~K or (b) $T_{eq}=1000$~K. Each curve corresponds to a different value of $L_{p}/M_p$: $10^{-12.5}~\rm{W\,kg^{-1}}$ (thin solid), $10^{-12.0}~\rm{W\,kg^{-1}}$ (thin dashed), $10^{-11.5}~\rm{W\,kg^{-1}}$ (thin dot-dashed), $10^{-11.0}~\rm{W\,kg^{-1}}$ (thin dotted), $10^{-10.5}~\rm{W\,kg^{-1}}$ (thick solid), $10^{-10.0}~\rm{W\,kg^{-1}}$ (thick dashed), and $10^{-9.5}~\rm{W\,kg^{-1}}$ (thick dot-dashed). Black lines denote our model radii (defined at a radial optical depth $\tau=2/3$), while the corresponding blue lines represent radii corrected for the transit radius effect. The green curves show the effective planet Roche-lobe radius for four different choices of host-star properties representative of spectral classes M5 V, M0 5V, K0 V, and G2 V  (in order of increasing Roche lobe radii). The K0 V and G2 V Roche-lobe radii are beyond the scale of the $T_{eq}=500$~K plot.}
\label{fig:MRTint}
\end{center}
\end{figure*}

We explored the parameter space of $M_{core}$, $M_{env}$, $T_{eq}$, and $T_{int}$ with our equilibrium model. Figures~\ref{fig:MRgmf} and \ref{fig:MRTint} present a selection of mass-radius ($M_p-R_p$) curves at (a) $T_{eq}=500$~K and (b) $T_{eq}=1000$~K. Figure~\ref{fig:MRgmf} displays the effect on the radius of varying the envelope mass fraction, while Figure~\ref{fig:MRTint} shows the effect of varying the planet's intrinsic luminosity, $L_p=4\pi R_p^2\sigma T_{int}^4$. The thick solid line is common between Figure~\ref{fig:MRgmf} and \ref{fig:MRTint}, representing $M_{env}=0.2M_p$ and $L_{p}/M_p=10^{-10.5}~\rm{W\,kg^{-1}}$. $L_{p}/M_p=10^{-10.5}~\rm{W\,kg^{-1}}$ corresponds to both the $8.3~M_{\oplus}$ evolution model (Run Ia) at 4~Gyr, and the $3.19~M_{\oplus}$ evolution model (Run IIa) at 1~Gyr (independent of $T_{eq}$). 

The $M_p-R_p$ curves for low-mass planets with voluminous gas layers show several notable features. First, the planet radii (at constant envelope mass fraction, $T_{eq}$, and $L_p/M_p$) increase dramatically toward low planet masses. This is due to the low surface gravities, and thus large atmospheric scale heights found at low masses. 
Second, the radius of planets having identical envelope mass fractions, $M_{env}/M_p$, are remarkably insensitive to the planet mass when $M_p\gtrsim 15~M_{\oplus}$. At these masses, increased compression of the envelope offsets the effect of increasing the planet mass.
Third, for planets of identical total mass (within the mass range plotted) the planet radius increases monotonically with the envelope mass fraction. 
Fourth, $T_{int}$ and $T_{eq}$ both have a stronger effect on the radius of low mass planets compared to their more massive counterparts. This is understandable, 
because, given the same envelope mass fraction, in lower mass planets the envelope accounts for a larger fraction of the planet radius. 

\begin{figure}[htb]
\epsscale{0.98}
%\epsscale{0.49}
\begin{center}
\plotone{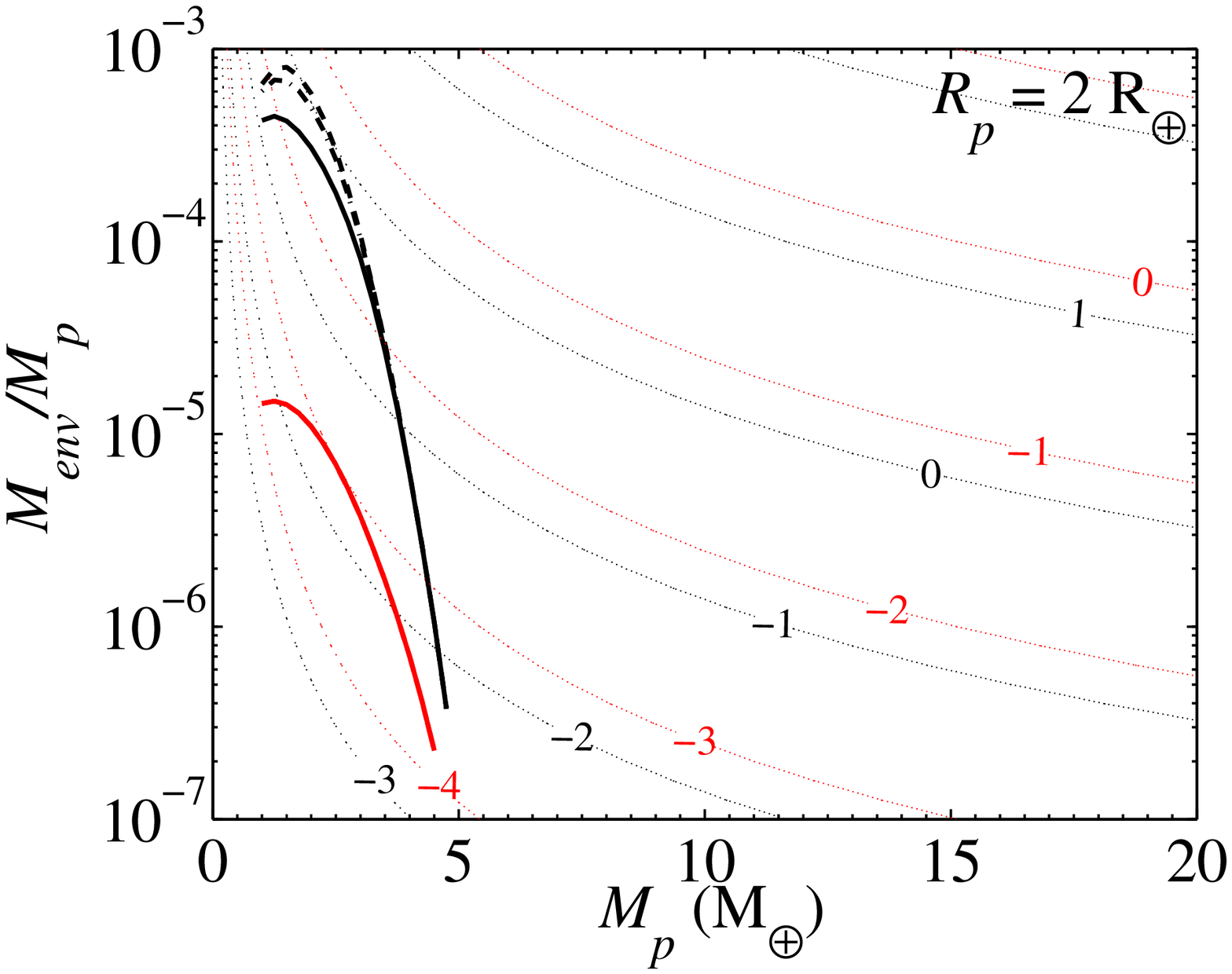}
\plotone{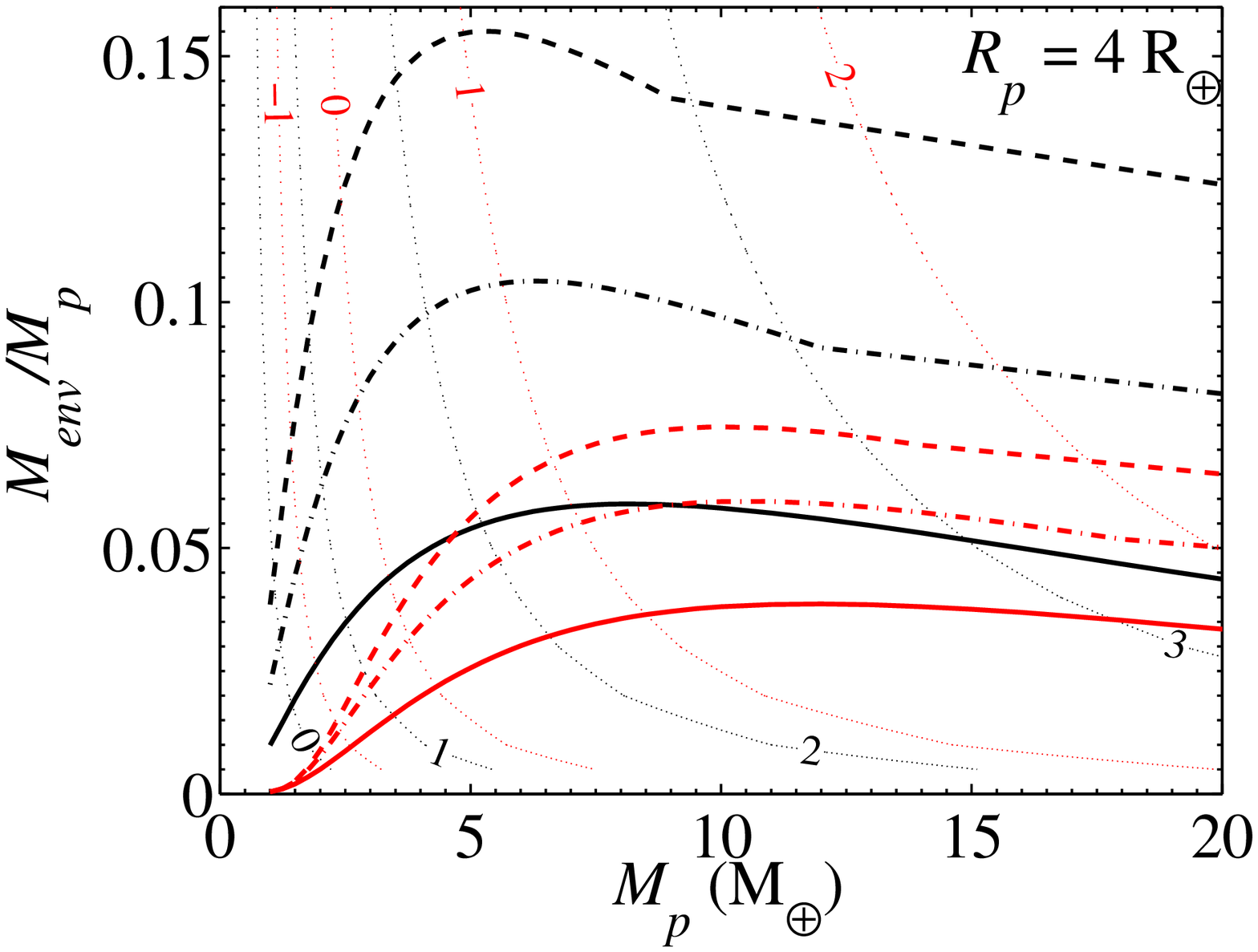}
\plotone{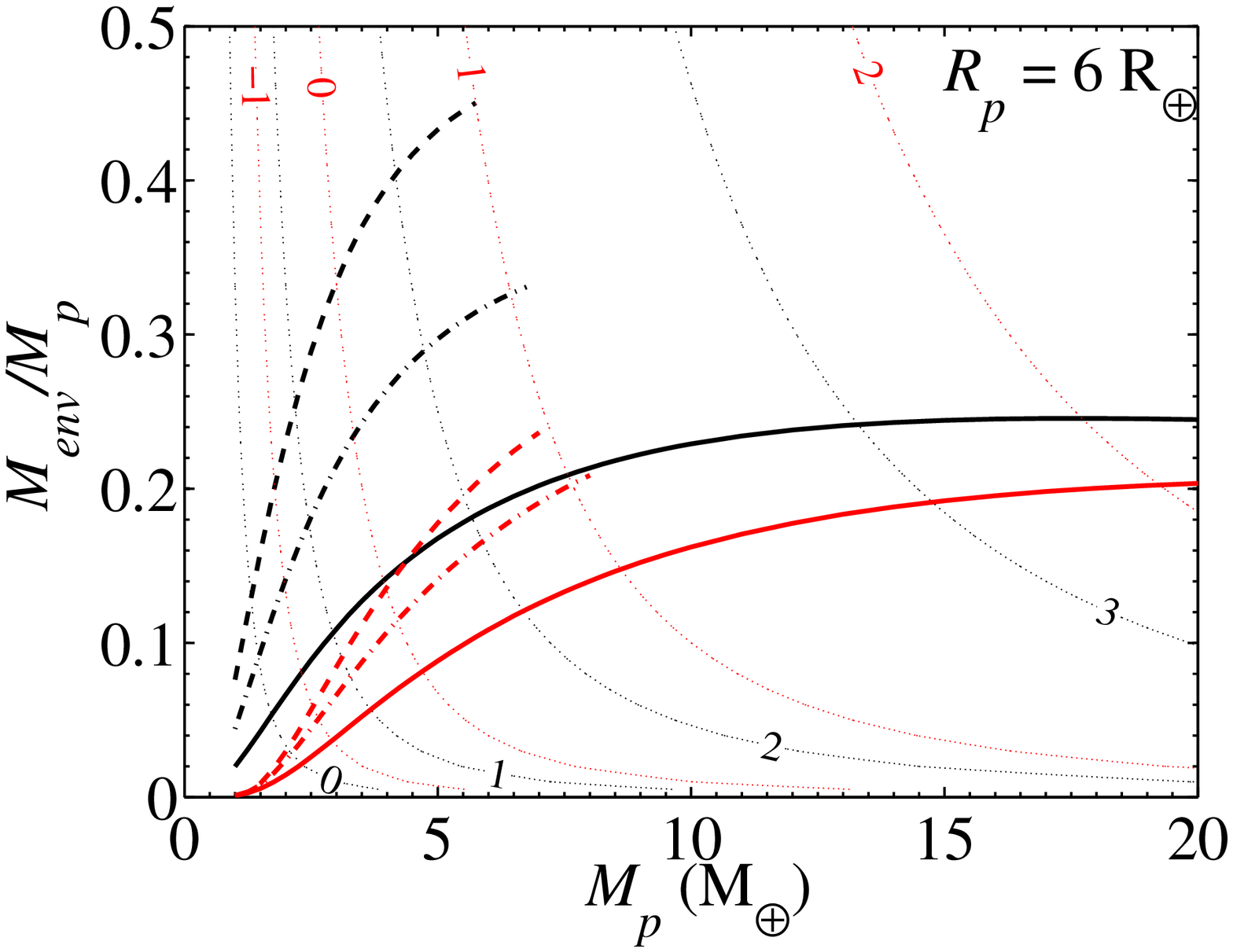}
%\plotone{lowrho2gG_wpv36hG_gmf_R1.eps}
%\plotone{lowrho2gG_wpv36hG_gmf_R2.eps}
%\plotone{lowrho2gG_wpv36hG_gmf_R3.eps}
\caption{Planet mass and envelope mass that are consistent with a particular planet radius, for planets comprised of ice-rock interiors surrounded by H$_2$ and He in protosolar proportions. These models represent planets that formed beyond the snow line by core nucleated accretion. We plot the envelope mass fraction as a function of total planet mass for planets with radii (a) $R_p=2~R_{\oplus}$, (b) $4~R_{\oplus}$, and (c) $6~R_{\oplus}$. Black curves represent planets at $T_{eq}=500$~K, while red curves correspond to $T_{eq}=1000$~K. The line style indicates the planet luminosity: $L_{p}/M_p=10^{-11}~\rm{W\,kg^{-1}}$ (dashed), $L_{p}/M_p=10^{-10}~\rm{W\,kg^{-1}}$ (dot-dash), and $L_{p}/M_p=10^{-9}~\rm{W\,kg^{-1}}$ (solid). The thin dotted lines are contours of constant envelope mass loss timescale,  $t_{\dot{M}}\equiv M_{env}/\dot{M}$. Each contour is labeled with $\log(t_{\dot{M}}/{\rm Gyr})$ for $\epsilon L_{XUV}/L_{BOL}=10^{-6}$, and can easily be scaled for other choices of $\epsilon L_{XUV}/L_{BOL}$ using Equation~(\ref{eq:mdot}).}
\label{fig:RParam}
\end{center}
\end{figure}

Planet radii between 2 and $6~R_{\oplus}$ are of special interest, because \textit{Kepler} is finding a large number of planet candidates within this size-range \citep{BoruckiEt2011ApJ, BoruckiEt2011AstroPh}. We plot in Figure~\ref{fig:RParam} combinations of $M_{env}$ and $M_p$ that yield planet radii within this range. Planets at $2~R_{\oplus}$ can contain at most 0.08\% of their mass in H/He at $T_{eq}=500$~K, and at most 0.0015\% at $T_{eq}=1000$~K. Larger planets can support more massive envelopes. A $6~R_{\oplus}$ planet at $T_{eq}=500$~K requires an envelope accounting for at least a few percent of the planet mass. At $T_{eq}=1000$~K and $6~R_{\oplus}$,  between 0.1\% and 23\% H/He by mass is possible, depending on the planet mass and intrinsic luminosity. 

It is important to note that the $M_p-R_p$ relations in Figures~\ref{fig:MRgmf}, \ref{fig:MRTint}, and \ref{fig:RParam} are not isochrons, but correspond instead to constant total intrinsic luminosity per unit mass, $L_{p}/M_p$. The total intrinsic luminosity, $L_p$, is the sum total of heating from radioactive decay, cooling of the planet core, and contraction of the planet envelope. In the evolution calculations from Table~\ref{tab:Rp}, the planet luminosity contribution from envelope contraction alone ranges from $10^{-9.8}$ to $10^{-11.2}~\rm{W\,kg^{-1}}$  at 1~Gyr and from $10^{-10.5}$ to $10^{-12.4}~\rm{W\,kg^{-1}}$ at 4~Gyr. Some of the low-$L_p$ curves in Figures~\ref{fig:MRTint} and \ref{fig:RParam} do not extend to higher masses because they encounter unphysically low planet interior entropies. Although $L_{p}$ is a proxy for the age of the planet, the relationship between $L_{p}$ and planet age depends on the planet's mass, composition,  abundance of radioactive isotopes, insolation history and dynamical history. Since our equilibrium models are presented at a specified $L_p$, we have side-stepped the issue of relating $L_p$ to planet age and present the model radii in a way such that they can be applied to many different evolution scenarios. Our aim with the equilibrium models is to broadly explore parameter space; it is beyond the scope of this work to relate $L_{p}$ and age directly by simulating all possible planet evolution histories.  

Simulated planet radii for planets at $T_{eq}=1000$~K may be in error by up to 20\%. The problem is in extrapolating the opacity tables at the high pressure end. This in turn makes the radiative-convective boundary uncertain (a deeper radiative-convective boundary makes for a smaller planet). Planets at $T_{eq}=500$~K are less affected by this opacity-caused radius problem ($\lesssim10\%$ radius uncertainty for $M_p\geq3~M_{\oplus}$). This issue affects both our equilibrium and evolution models.

\section{Planet Formation by Outgassing of Hydrogen}
\label{sec:outgas}

\subsection{Model}

Outgassing provides a mechanism for low-mass terrestrial planets to acquire an atmosphere even if they fail to accrete H and He from the protoplanetary nebula. In this section we explore the optimum conditions for a planet to acquire a voluminous gas envelope through outgassing. We base our model approach on \citet{ElkinsTanton&Seager2008aApJ, ElkinsTanton&Seager2008bApJ}, with the improvements of a more detailed interior structure model and a calculation of the planet radius.

We focus on outgassing of H$_2$ produced when water reacts with metallic Fe in accreting materials during planet formation \citep{Ringwood1979book, Wanke&Dreibus1994RSPTA, ElkinsTanton&Seager2008aApJ}. Hydrogen gas has the potential to yield the most voluminous outgassed atmospheres, being both of low-molecular weight and (for some planetesimal compositions) degassed in substantial quantities. Although we do not consider these processes in detail here, in general, outgassing may also proceed during accretion as impinging planetesimals are heated and vaporized upon impact; during magma ocean solidification as volatiles are partitioned between the atmosphere and melt; and during volcanic/tectonic activity after the planet has formed. 

The reaction between water and metals during planetary accretion and differentiation intrinsically links the planet's interior structure to its initial atmosphere's mass and composition. Metallic iron forming the planet will either differentiate to contribute to the planet iron core, or become oxidized and incorporate into the planet mantle. Given an initial composition for the primordial material forming a planet, there are two extremes to the eventual planet outcomes. If none of the available water and metals in the accreting materials react (reducing conditions), the planet will have a maximally massive metallic core, relatively iron-poor mantle, minimal outgassed H$_2$, and maximal leftover H$_2$O. In contrast, if the water and metals react to the maximal extent possible (oxidizing conditions), the planet will have a minimal iron core mass, iron-rich mantle, maximal outgassed H$_2$, and minimal leftover H$_2$O. When Fe is the limiting reagent, this extreme will correspond to a coreless planet \citep{ElkinsTanton&Seager2008bApJ}.

To bound the radii of outgassed rocky planets, we consider the end-member case of a planet formed purely from high iron enstatite (EH) primordial material. The motivation for this choice is three-fold. First, out of all meteoritic compositions, EH material has the potential to degas the most H$_2$ per unit mass \citep[up to 3.6\%,][]{ElkinsTanton&Seager2008aApJ}. Second, the oxygen isotope mixing model \citep{Lodders2000SSRv} predicts that the Earth accreted from material that was 70\% EH chondritic matter by mass. Third, heating of EH material releases a  low mean molecular weight atmosphere; \citet{Schaefer&Fegley2010bIcarus} calculated 44\% H$_2$, 31\% CO, 17\% H$_2$O, 5\% CO$_2$, and 3\% other molecules by volume under their nominal conditions (1500~K, 100 bars).
Thus, complete oxidation of an EH planet should achieve effectively the maximum radius plausible for planets with outgassed atmospheres.

For the EH material we adopt the chemical composition of meteorite ALHA77295 from \citet{Jarosewich1990Meteoritics}. We distill the mineralogy in our model to include only the most plentiful and important constituents: metallic Fe, FeS, FeO, Fe$_2$O$_3$, MgO, SiO$_2$, H$_2$O, and H$_2$. Following an approach similar to \citet{SotinEt2007Icarus}, less abundant elements are represented by their most similar neighbors in the periodic table: metallic Ni is added to metallic Fe, Ca is added to Mg, and Al is divided equally (by number) between Si and Mg to preserve charge conservation. Other trace constituents (TiO$_2$, Cr$_2$O$_3$, MnO, Na$_2$O, K$_2$O, P$_2$O$_5$, Co, which combined account for less than 2.2\% by mass) are neglected. The resulting simplified composition adopted for the primordial rocky EH planetesimals consists of (by mass) 38.2\% SiO$_2$, 25.2\% metallic Fe, 14.3\% FeS, 20.6\% MgO, 1.7\% H$_2$O. Note that  H$_2$O included in the EH material is adsorbed to the surface or chemically bound to the minerals.

We consider planets initially formed from a mixture of EH material and H$_2$O ice. The H$_2$O ice is in addition to the 1.7\%  H$_2$O by mass included in the EH minerals. 
We compute the planet bulk composition after outgassing from stoichiometry (Table~\ref{tab:Outcomp}), assuming some fraction of the accreted iron reacted with water (Fe~+~H$_2$O~$\rightarrow$~FeO~+~H$_2$) before sinking to form the planet's metallic core.
We note that although we consider only Fe in our reduced EH chemical composition, Ni can also form oxides and be incorporated in silicates. Nickel accounts for 8\% of the generalized metallic Fe in our distilled EH chemical composition -- the Ni abundance is abundance in ALHA77295 is 1.83\% by mass.
We do not vary the S mass fraction of the iron core in our models, effectively assuming  metallic Fe and FeS oxidize in equal proportions. We do not follow any S released in the conversion of FeS to FeO. 

Our interior models of outgassed planets comprise up to four chemically distinct layers: an Fe/FeS core, silicate mantle, water layer, and hydrogen atmosphere. The bulk chemical composition of the planet after outgassing determines the relative masses of the planet layers and the composition of the silicate mantle. All of the degassed H$_2$ is included in a gas layer surrounding the planet. We place all of the FeS  and metallic iron in the planet core. We model the H$_2$O in a differentiated water layer surrounding the mantle, although in practice some water may be sequestered into the silicates \citep[e.g.,][ and references therein]{ElkinsTanton2008E&PSL}. All of the remaining species (SiO$_2$, MgO, FeO, Fe$_2$O$_3$) make up the mantle. The ratio of MgO/FeO sets the Mg \# of the silicates (Mg \# = Mg/(Mg+Fe) by number). We adjust the mantle equation of state to reflect the relative abundances of SiO$_2$, MgO, FeO and Fe$_2$O$_3$, modeling the silicates as a mixture of (Mg,Fe)O  magnesiowustite \citep[][]{ElkinsTanton2008E&PSL}, Fe$_2$O$_3$ hematite \citep{Wilburn&Bassett1978AmMineral}, and stishovite SiO$_2$ \citep{AndraultEt1998Sci}. Outgassed bulk compositions and the corresponding planet properties are reported in Table~\ref{tab:Outcomp}. 

\begin{table*}[thb]
 \caption{Bulk compositions of EH-composition planets with outgassed H$_2$ envelopes. The first column represents the fraction of accreted iron that is oxidized and incorporated in the planet's mantle. The next four columns give the composition of the EH planet after outgassing, assuming all outgassed H$_2$ is retained. Negative entries in the H$_2$O column indicate a water deficit and represent the proportion of additional water (beyond what is included in the EH material) that needs to be accreted in order to oxidize the specified fraction of iron. The last five columns represent the chemical make-up of the silicate mantle, and determine the mantle equation of state. In rows 2--4 we neglect Fe$_2$O$_3$ and assume only FeO is produced when iron is oxidized. Row 5 lists the extreme end-member case where all iron is oxidized to Fe$_2$O$_3$.}
 \label{tab:Outcomp}
 \centering
 \vspace*{1ex}%
 \resizebox{1.0\textwidth}{!}{%
 \small
 \begin{tabular}{l|cccc|ccccc}
 \hline
 &&&&&&&&\\
\% Fe oxidized & Core wt \% & Silicate wt \% & H$_2$O wt \% excess &  H$_2$ wt \%  & \multicolumn{4}{c}{Silicate Composition}\\
&&&&& MgO wt \%& FeO wt \% & Fe$_2$O$_3$ wt \% & SiO$_2$ wt \%& Mg \#\\
 \hline\hline
 0.0 & 39.5 & 58.8 & 1.7 & 0.0 & 35.1 & 0.0 & 0.0 & 64.9 & 1.00 \\
 15.2 & 33.8 & 66.0 & 0.0 & 0.2 & 31.5 & 10.3 & 0.0 & 58.3 & 0.85 \\
 50.0 & 19.5 & 79.9 & -3.7 & 0.6 & 25.5 & 27.3 & 0.0 & 47.2 & 0.62 \\
 100.0 & 0.0 & 98.8 & -8.6 & 1.2 & 20.0 & 42.9 & 0.0 & 37.1 & 0.45 \\ 
  100.0 & 0.0 & 98.3 & -13.0 & 1.7 & 19.1 & 0.0 & 45.5 & 35.4 & 0.45 \\ 
 \hline
 \end{tabular}
 }
\end{table*}

\subsection{Results}
\label{sec:outgasres}

We find that planets accreted from solid bodies that were abundant in our solar nebula can degas at most 1.7\% of their mass in H$_2$. This limit obtains for a fully-degassed coreless EH composition planet that accreted just enough additional water (13.0\% by mass) to fully oxidize all available iron to Fe$_2$O$_3$.  EH material alone does not contain sufficient H$_2$O on its own to oxidize all the metallic Fe within its bulk (only up to 15.2\% of the Fe). The accreted material must include an additional 8.6\% H$_2$O by mass in order to convert all the metallic Fe into FeO, or an additional 13.0\% H$_2$O by mass to convert all the metallic Fe into Fe$_2$O$_3$. With any more water than this, the metallic Fe becomes the limiting reagent. The maximal outgassed H$_2$ atmosphere that we derive here is slightly lower than the value 3.6 wt \% H$_2$ found by \citet{ElkinsTanton&Seager2008aApJ}. Differences in the representative EH chemical compositions assumed account for this disparity.

\begin{figure*}[htb]
\begin{center}
\epsscale{1}
\plottwo{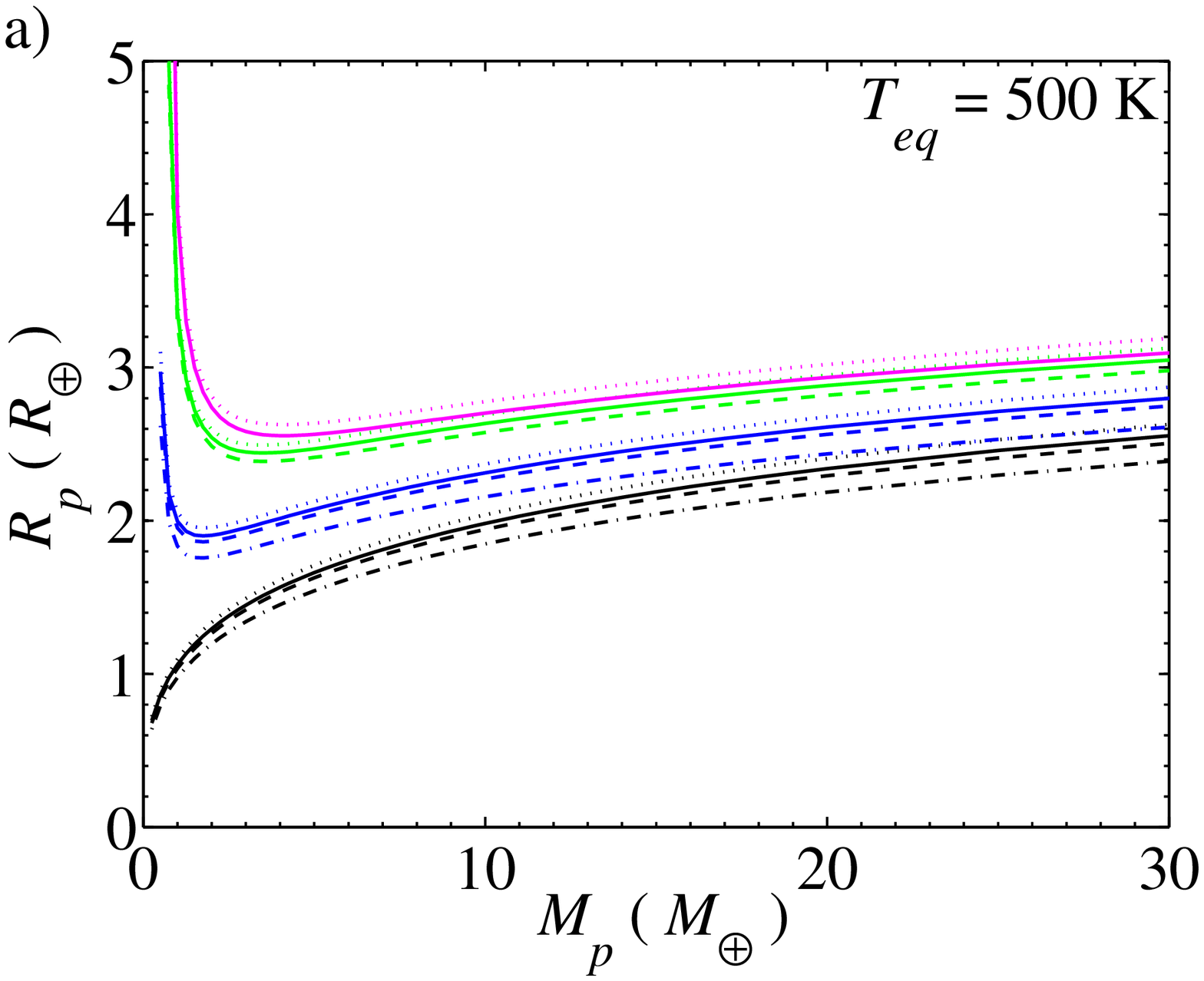}{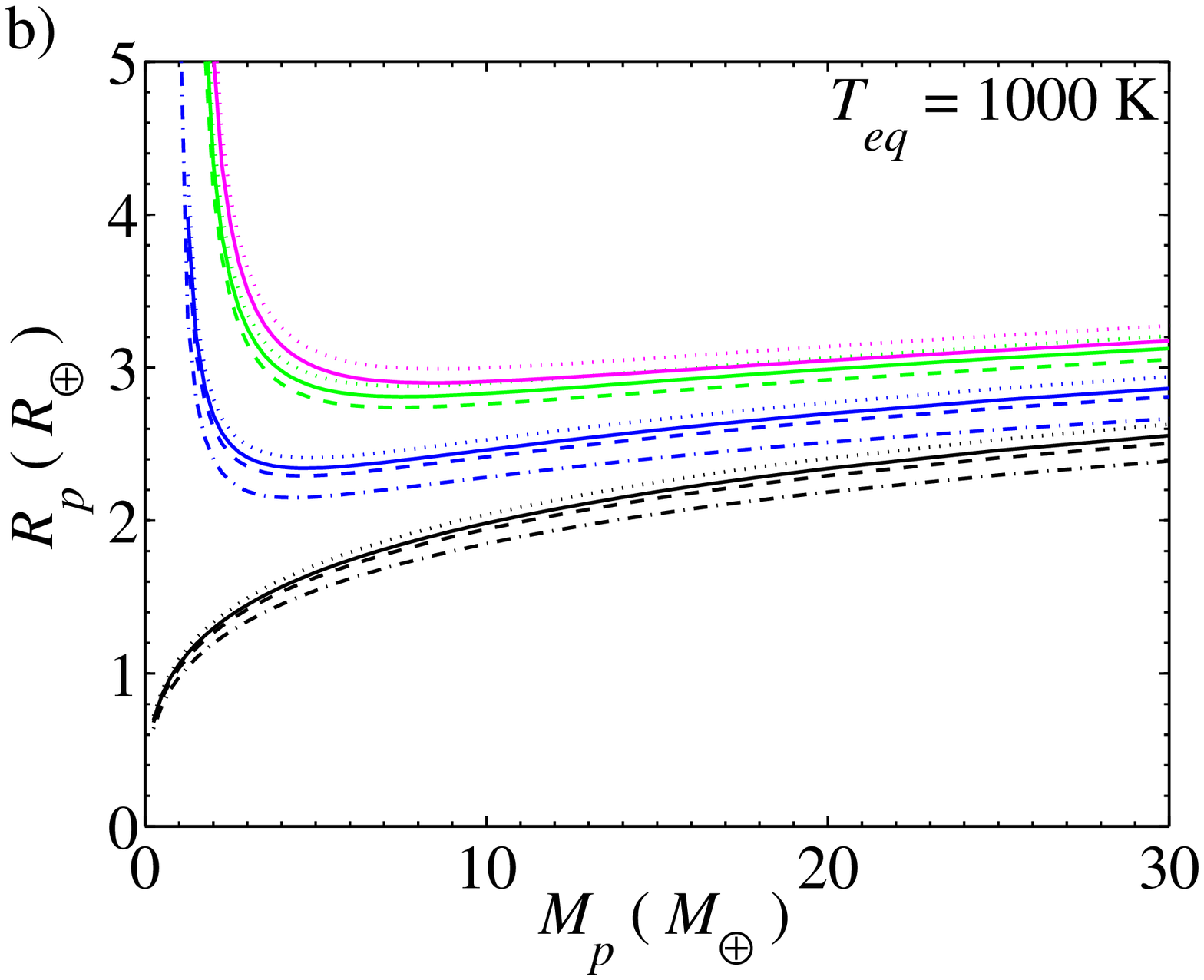}
%\plottwo{wpv36hR_EHOutgassing_Teq500LoverM105.eps}{wpv36hR_EHOutgassing_Teq1000LoverM105.eps}
\caption{Mass-radius relations for exoplanets with outgassed H$_2$ envelopes. The planets are assumed to have formed purely from a combination of EH chondrite material and water ice. Accreting material with 20\% water ice by mass (dotted lines), 13\% water ice by mass (solid lines) 8.6\% water ice by mass (dashed lines), and no additional water ice (dot-dashed lines) are considered. The line color indicates the fraction of accreted iron that reacted with water. Black corresponds to planets with no outgassed H$_2$ and a maximally massive iron core (0\% Fe reacted). Blue corresponds to planets where 15.2\% of the Fe reacted - the maximum amount possible for pure EH material without added water. Green represents an end-member case wherein all the metallic Fe that accreted to the planet is converted to FeO. Finally, magenta lines correspond to planets that outgassed the maximum possible H$_2$ for their initial chemical makeup -- 100\% of their accreted iron is oxidized to Fe$_2$O$_3$. Both the green and magenta $M_p-R_p$ relations represent core-less planets, but they differ in the oxidation state of iron inside the planet (FeO versus Fe$_2$O$_3$) and in the overall proportion of H$_2$ released. Planet equilibrium temperatures of a) $T_{eq}=500$~K and b) $T_{eq}=1000$~K are shown. A fiducial intrinsic luminosity $L_p/M_p=10^{-10.5}~\rm{W\,kg^{-1}}$ is assumed in all cases. These curves do not include atmospheric escape of H$_2$. }
\label{fig:outgasR}
\end{center}
\end{figure*}

Mass-radius relations for planets harboring H$_2$-envelopes from outgassing are shown in Figure~\ref{fig:outgasR} at both $T_{eq}=500$~K (Figure~\ref{fig:outgasR}a) and $T_{eq}=1000$~K (Figure~\ref{fig:outgasR}b). The blue dot-dashed curve provides an upper limit on the radius of planets accreted from primordial chondritic material alone (without additional water ice), corresponding to the extreme where the oxidizing reaction proceeds until all of the H$_2$O bound to the minerals is expended and 0.2\% of the planet mass is released in H$_2$. After accreting enough additional water (13\% by mass) to convert all available Fe to Fe$_2$O$_3$, the magenta solid line represents planets having the maximal fraction of their mass (1.7\%) in a degassed H$_2$ envelope. This curve may be taken to bound the maximum radius/minimum density relation for planets with de-gassed H$_2$ envelopes, but no free H$_2$O. 

Planets that accreted more than 13.0\% by mass water with the EH chondrite material would have water left over even if all the metals in the planet iron core were expended in the outgassing reaction. In Figure~\ref{fig:outgasR} we show $M_p-R_p$ relations of an example with initially 20\% by mass water ice in the primordial composition (dotted curves). The fully degassed planets with excess water have, in fact, a lower average density compared to the planets with the highest mass fraction of de-gassed H$_2$ -- the effect of the lower density ice-rock interior offsets the decreased proportion of H$_2$. In Figure~\ref{fig:outgasR}, we model the H$_2$O layer as a distinct chemical layer below the outgassed H$_2$ envelope, but mixing of H$_2$O and H$_2$ is another possibility. If H$_2$O and H$_2$ are mixed in the envelope, the planet radii would be smaller than the model radii in Figure~\ref{fig:outgasR} due to the decreased atmospheric scale height compared to the differentiated case.

The radii of the outgassed planets depend on the intrinsic luminosity of the planet. In Figure~\ref{fig:outgasR}, we show mass-radius relations for planets with $L_p/M_p=10^{-10.5}~\rm{W\,kg^{-1}}$. 
 Increasing (decreasing) the planet's intrinsic luminosity by a factor of 10 affects the planet radii in Figure~\ref{fig:outgasR} by at most +16\% (--9.5\%) at $5~M_{\oplus}$ and +4.5\% (--3.2\%) at $30~M_{\oplus}$. Small planet masses and high H$_2$ contents both increase the radius dependence on $L_p$.

\begin{figure}[htb]
\epsscale{0.98}
%\epsscale{0.49}
\begin{center}
\plotone{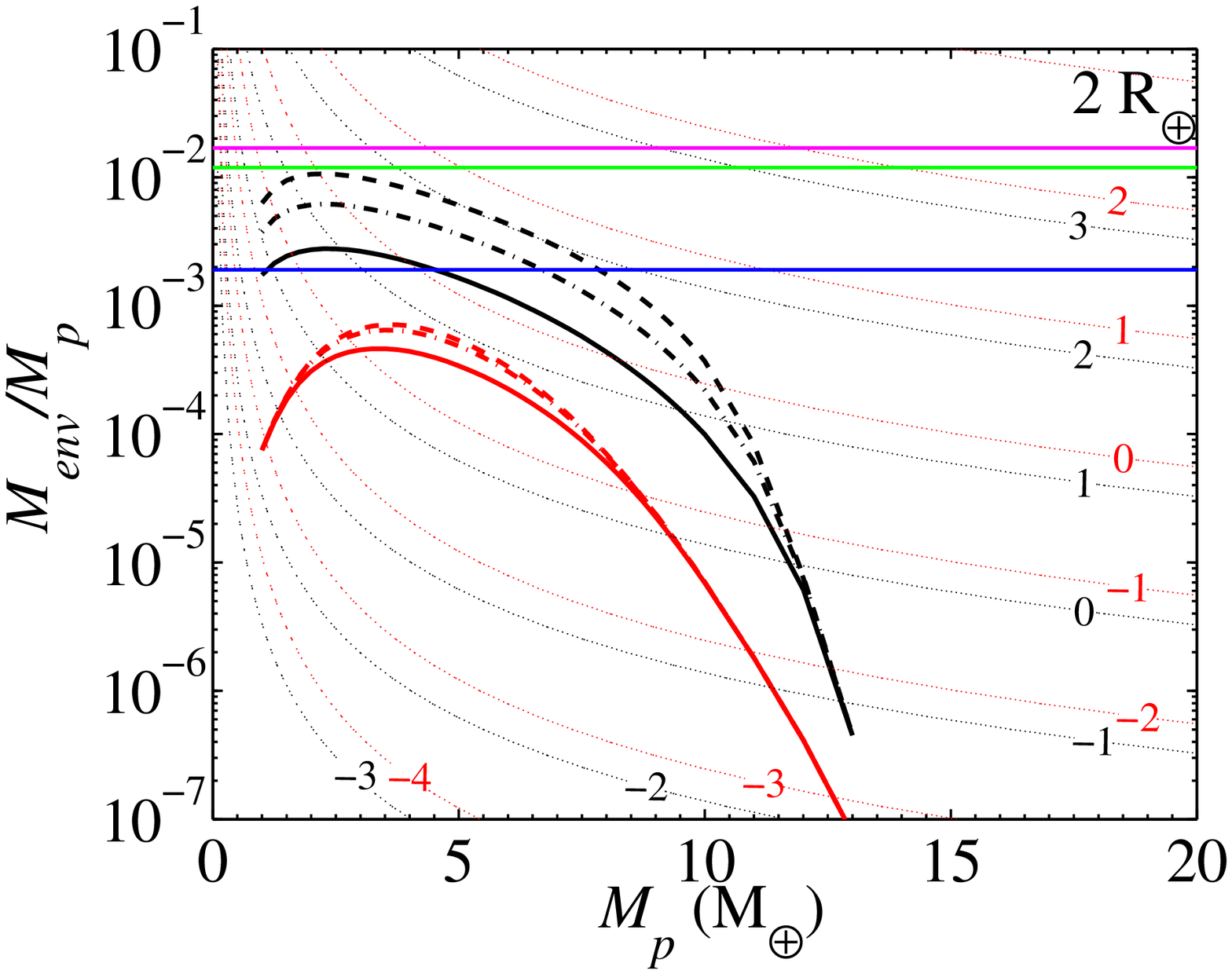}
\plotone{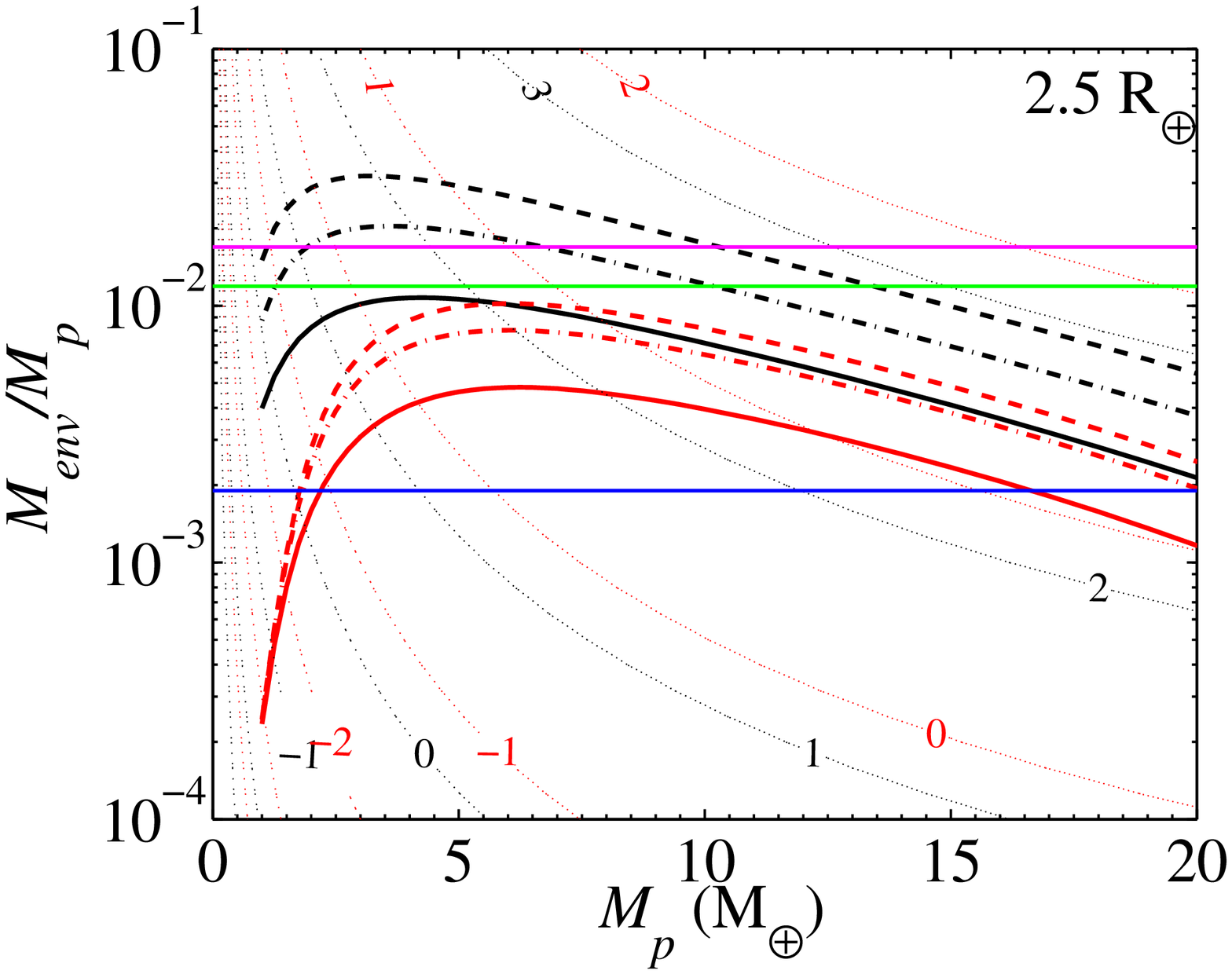}
\plotone{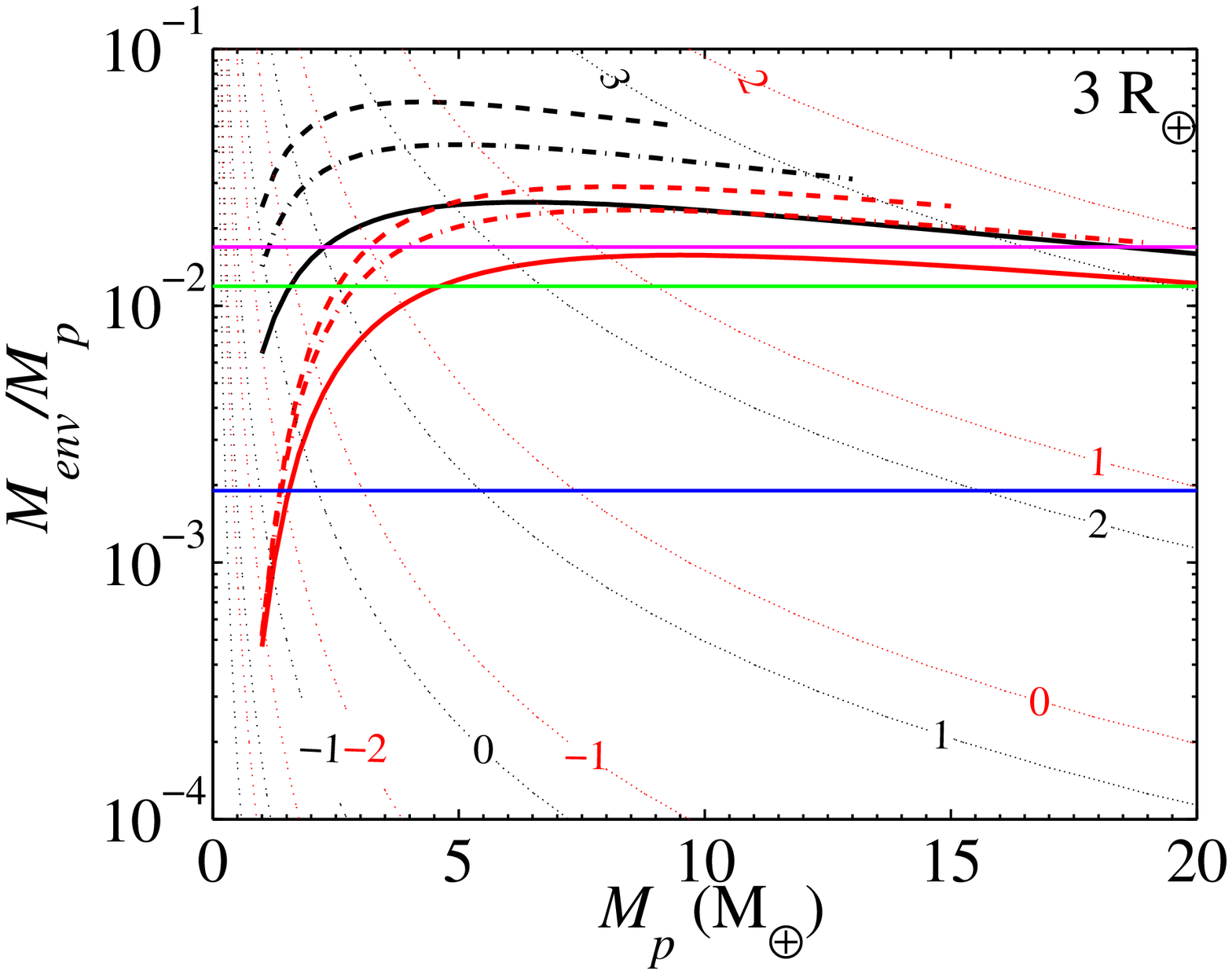}
%\plotone{wpv36hG_EHOutgassing_gmf_R1.eps}
%\plotone{wpv36hG_EHOutgassing_gmf_R2.eps}
%\plotone{wpv36hG_EHOutgassing_gmf_R3.eps}
\caption{Planet mass and outgassed H$_2$ envelope mass that are consistent with a particular planet radius, for EH-composition planets without H$_2$O on their surface or in their envelopes.  We plot the envelope mass fraction as a function of the total mass of the planet for planets with radii (a) $R_p=2~R_{\oplus}$, (b) $2.5~R_{\oplus}$, and (c) $3~R_{\oplus}$. 
Horizontal lines indicate the maximal H$_2$ wt \% degassed in three limiting cases: if all H$_2$O adsorbed in the EH material reacts with metals (0.2\%, blue), if all Fe in the EH material is converted to FeO (1.2\%, green), and if all Fe in the EH material is converted to Fe$_2$O$_3$ (1.7\%, magenta). 
This figure is the outgassing analog to Figure~\ref{fig:RParam} for core nucleated accretion, and all the red and black lines follow the same naming conventions. Black curves represent planets at $T_{eq}=500$~K, while red curves correspond to $T_{eq}=1000$~K. The line style indicates the planet luminosity: $L_{p}/M_p=10^{-11}~\rm{W\,kg^{-1}}$ (dashed), $L_{p}/M_p=10^{-10}~\rm{W\,kg^{-1}}$ (dot-dash), and $L_{p}/M_p=10^{-9}~\rm{W\,kg^{-1}}$ (solid). The thin dotted lines are contours of constant envelope mass loss timescale,  $t_{\dot{M}}\equiv M_{env}/\dot{M}$. Each contour is labeled with $\log(t_{\dot{M}}/{\rm Gyr})$ given $\epsilon L_{XUV}/L_{BOL}=10^{-6}$. }
\label{fig:outRParam}
\end{center}
\end{figure}

We explore in Figure~\ref{fig:outRParam} the mass of H$_2$ required by EH composition planets to reach radii of 2 to 3~$R_{\oplus}$. Figure~\ref{fig:outRParam} is the outgassing analog to Figure~\ref{fig:RParam} for core nucleated accretion. In Figure~\ref{fig:outRParam}, we restrict our attention to planets without significant amounts of H$_2$O on their surface or in their envelopes. The envelope mass fractions, $M_{env}/M_p$, at a specified radius are not strongly sensitive to the distribution of Fe within the planet interior (i.e., whether the Fe is differentiated in the metallic core, or included in the mantle as oxides) -- we show the case where all Fe is oxidized to FeO. Upper bounds on the H$_2$ wt \% for several of the limiting cases in Table~\ref{tab:Outcomp} are indicated by colored horizontal lines. 

Our main conclusion from this section is that planets of mass $M_p<30~M_{\oplus}$ with outgassed H atmospheres typically have radii less than $3~R_{\oplus}$ (Figures~\ref{fig:outgasR} and \ref{fig:outRParam} ). Larger radii are found at the low-mass extreme of the $M_p-R_p$ relations in Figure~\ref{fig:outgasR}, but correspond to planets with very tenuous, loosely bound, envelopes. Outgassing of H$_2$ from planets accreted from rocky material alone most likely cannot account for the {\it Kepler} planet candidates with radii between 3 and 6~$R_{\oplus}$.  

\section{Mass Loss from Low-Density Envelopes}
\label{sec:mdot}

A major question is whether the high $T_{eq}$, light element, low gravitational binding energy envelopes modeled above are stable and could be retained over gigayear timescales. It is precisely in the low-mass, low-molecular weight, high $T_{eq}$ regime we are considering in which planets are expected to be most susceptible to mass loss. Below we consider, in turn, the importance of 
Roche-lobe overflow, and XUV-driven atmospheric escape.
 
Roche-lobe overflow can limit the radii of low-density planets at close orbital separations from their host stars. Our planet interior model assumes spherical symmetry and neglects tidal forces, but this approximation starts to break down for planets near their star. 
 The effective radius of a planet's Roche lobe is approximated by
 \begin{equation}
 \frac{r_L}{a} = \frac{0.49q^{2/3}}{0.6q^{2/3}+\ln\left(1+q^{1/3}\right)}\approx 0.49q^{1/3} - 0.049q^{2/3}
 \end{equation}
 \noindent where $q\equiv M_p/M_\star$ \citep{Eggleton1983ApJL}. The Roche-lobe radius sets a firm upper limit on the planet radius; any material outside the planet's Roche lobe is not gravitationally bound to the planet and can escape. We plot planet Roche-lobe radii in Figures~\ref{fig:MRgmf} and ~\ref{fig:MRTint} for a sampling of representative host star properties:  G2 ($1~M_{\odot}$, $1~L_{\odot}$), K0 ($0.79~M_{\odot}$, $0.552~L_{\odot}$), M0 ($0.51~M_{\odot}$, $0.077~L_{\odot}$), M5 ($0.21~M_{\odot}$, $0.0076~L_{\odot}$) \citep{Carroll&Ostlie2006book}. In computing the Roche-lobe radii, we have assumed a planetary albedo $A=0$ when relating $T_{eq}$ to the semi-major axis, $a$; taking reflection into account with $A\neq0$ will result in smaller semi-major axes and smaller $r_L$.
Roche-lobe overflow is not an issue for $T_{eq}=500-1000$~K planets surrounding a solar analog star. 
In contrast, when orbiting an M star many of our low-density low-mass planets do fill their Roche lobes. 
Our equilibrium planet models are not a priori pegged to a given star spectral type. Tidal effects and the Roche-lobe radius set a lower bound on $M_\star$ for which the low $M_p$ tail of our equilibrium models are applicable.

XUV-driven mass loss is expected to be very important for low mass, low density planets.
This results from the combined effect of large cross-sections to stellar irradiation, low surface gravities, and low envelope binding energies. Predictions for the exoplanet mass loss rates suffer from unknowns in the stellar XUV fluxes, the conditions at the planet exosphere, and the mass loss efficiency. 
We consider energy-limited mass loss, \citep[e.g.,][]{LammerEt2003ApJ, LecavelierDesEtangs2007A&A, ValenciaEt2010A&A},
\begin{equation}
\dot{M}=-\frac{\epsilon\pi F_{XUV}R_{XUV}^2R_p}{GM_pK_{tide}}.
\label{eq:mdot}
\end{equation}
\noindent The efficiency $\epsilon$ represents the fraction of the energy in XUV photons incident on the planet that goes into unbinding particles in the planet atmosphere; we take $\epsilon=0.1$, but $\dot{M}$ can easily be rescaled to another choice of $\epsilon$. $F_{XUV}$ represents the flux of photoionizing radiation impinging on the planet. $K_{tide}$ is a correction factor that accounts for tidal effects in the Roche potential of planets in close proximity to their star \citep[given by equation 17 in ][]{ErkaevEt2007A&A}. Finally, $R_{XUV}$ reflects the planet radius at which XUV photons are absorbed. We estimate $R_{XUV}$ following order-of-magnitude arguments gleaned from Section 2 of \citet[][]{MurrayClayEt2009ApJ}, 
\begin{equation}
R_{XUV}\approx R_p+H_R\ln\left(\frac{P_RR_{XUV}^2}{N_Hm_HGM_p}\right),
\label{eq:r1}
\end{equation}
\noindent where $N_H\sim5\times10^{21}~\rm{m^{-2}}$ is roughly the column of neutral hydrogen needed to reach $\tau_{XUV}\sim1$, $P_R$ is the pressure at $R_p$, and $H_R$ is the pressure scale height at $R_p$ (where $\tau\sim1$ for visible light).  

\begin{figure*}[htb]
\begin{center}
\epsscale{1}
\plottwo{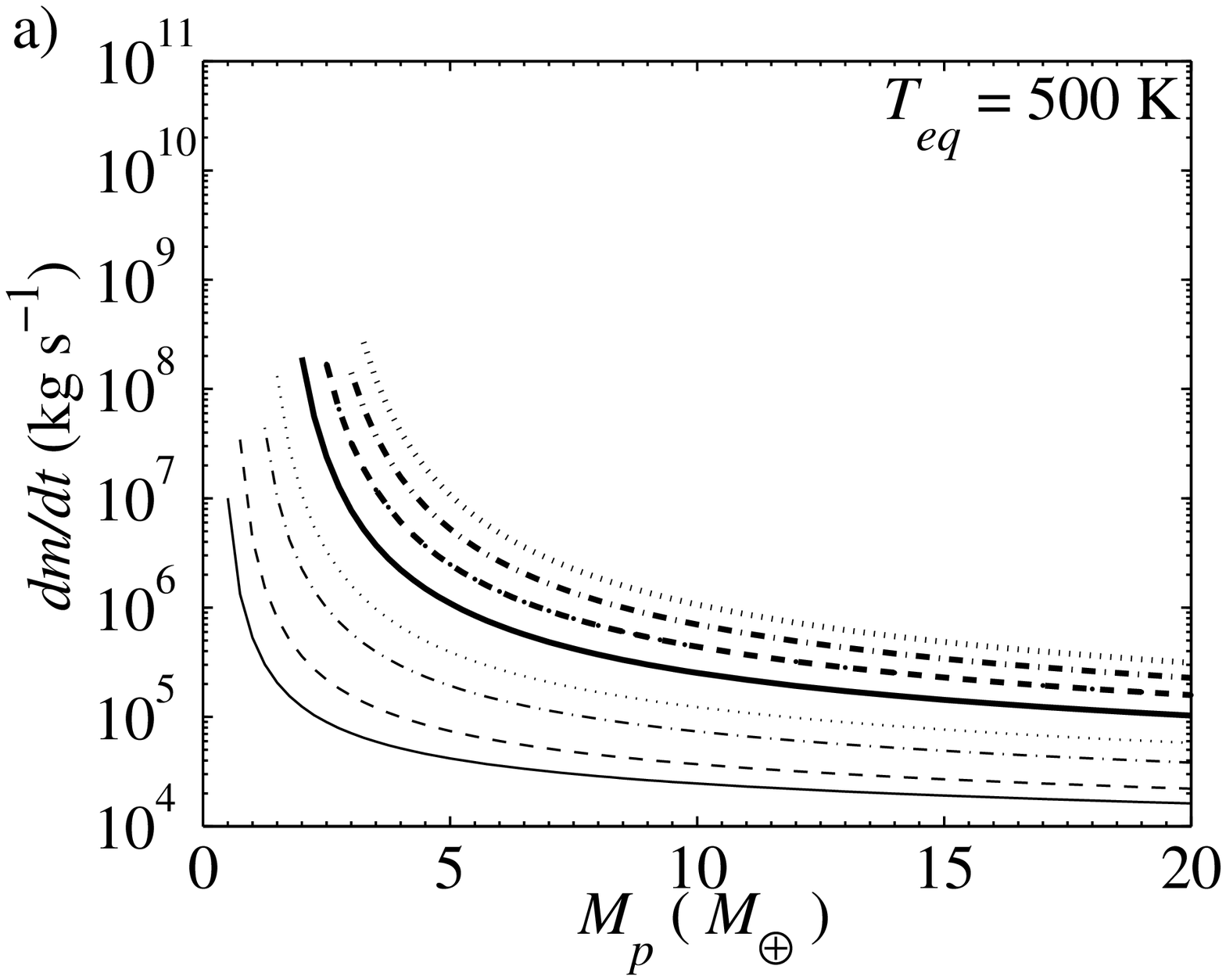}{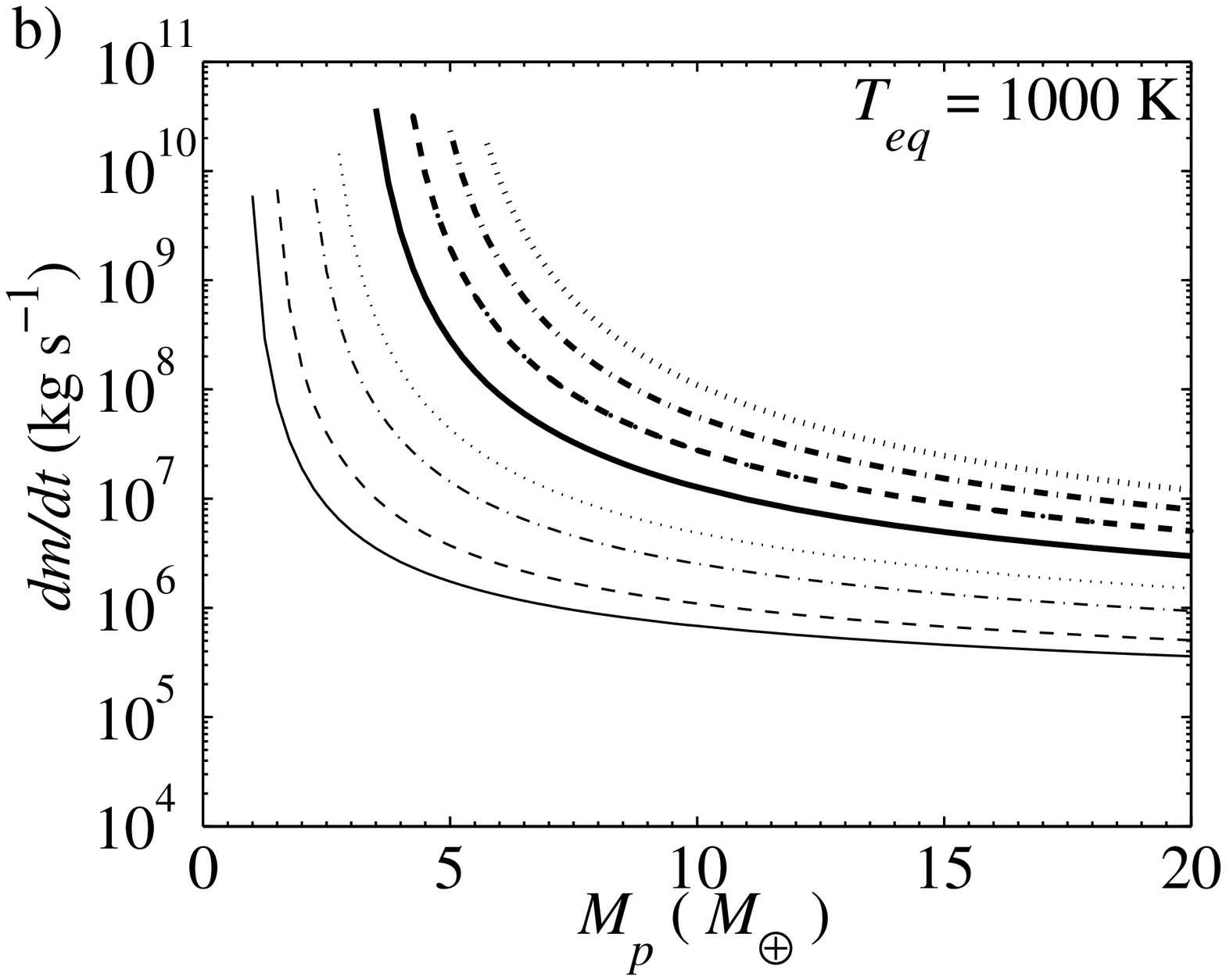}
%\plottwo{lowrho2gGice_wpv36h_Teq500_varygmf_mdot.eps}{lowrho2gGice_wpv36h_Teq1000_varygmf_mdot.eps}
\caption{Energy limited mass loss rates for the planet models in Figure~\ref{fig:MRgmf}. Mass loss rates are estimated for the case where the planets orbit a star with similar properties to our sun ($M_\star=1~M_{\odot}$, $L_\star=1~L_{\odot}$, and $L_{XUV}=3.4\times10^{-6}~L_{\odot}$). A mass loss efficiency of $\epsilon=0.1$ is assumed. The line styles have the same meanings and correspond to the same model planets as in Figure~\ref{fig:MRgmf}. Each curve corresponds to a different value of $M_{env}/M_p$: 0.001 (thin solid), 0.01 (thin dashed), 0.05 (thin dot-dashed), 0.1 (thin dotted), 0.2 (thick solid), 0.3 (thick dashed), 0.4 (thick dot-dashed), and 0.5 (thick dotted). All data in this plot have $L_{p}/M_p=10^{-10.5}~\rm{W\,kg^{-1}}$, and (a) $T_{eq}=500$~K or (b) $T_{eq}=1000$~K.}
\label{fig:mdot}
\end{center}
\end{figure*}

We take an illustrative example of planets orbiting a solar analog star to explore the order of magnitude of mass loss rates. Figure~\ref{fig:mdot} shows estimated mass loss rates for the planet models presented in Figure~\ref{fig:MRgmf}. For our assumed solar-twin host star, we compute $F_{XUV}$ for $T_{eq}$ by scaling the integrated solar XUV flux measured by \citet{RibasEt2005ApJ} ($F_{XUV\odot}=4.6\times10^{-3}~\rm{W\,m^{-2}}$ at 1~AU). We find that, for $L_{XUV}/L_{BOL}=3.4\times10^{-6}=L_{XUV\odot}/L_{BOL\odot}$ and $\epsilon=0.1$, planets at the low mass extreme of our $M_p-R_p$ relations have implausibly short envelope mass-loss timescales $t_{\dot{M}}\equiv M_{env}/\dot{M}\lesssim 1$~Gyr. 

We use energy limited mass loss (Equation~\ref{eq:mdot}) to include contours of constant $\log(t_{\dot{M}}/{\rm Gyr})$ in Figures~\ref{fig:RParam} and \ref{fig:outRParam}. The contour values represent $\log(t_{\dot{M}}/{\rm Gyr})$ corresponding to $(\epsilon L_{XUV}/L_{BOL}=10^{-6})$, but can easily be scaled to reflect other parameter choices:
\begin{equation}
t_{\dot{M}}\propto \left(\epsilon L_{XUV}/L_{BOL}\right)^{-1}.
\end{equation}
\begin{equation}
t_{\dot{M}}=\frac{M_{env}}{\dot{M}_{p}}
\end{equation}
At a specified $T_{eq}$, the $t_{\dot{M}}$ contours are independent of the host star mass so long as tidal effects can be neglected $(K_{tide}\approx1)$. For the $(T_{eq},\, R_p)$ combinations sampled in Figure~\ref{fig:RParam} and \ref{fig:outRParam}, this approximation holds for main sequence host stars that are K0 V or earlier, but breaks down for M stars. We emphasize that $t_{\dot{M}}$ gives an instantaneous measure of the time that the planet would take to lose its envelope at the calculated current mass loss rate. $\dot{M}$ is expected to vary over a planet's lifetime. Stars that are more active (e.g., younger) than our Sun would have higher photoionizing fluxes. 

We find that planets at the low mass extremes of Figures~\ref{fig:RParam} and \ref{fig:outRParam} have short envelope mass-loss timescales $t_{\dot{M}}\equiv M_{env}/\dot{M}\lesssim 1$~Gyr (assuming $\epsilon L_{XUV}/L_{BOL}=10^{-6}$). One could conceivably choose a threshold envelope loss timescale $t_{\dot{M}0}$ and then derive a lower bound on the planet mass at a given radius based on that assumption. We elaborate this possibility further in Section~\ref{sec:mmin}. 

\section{Discussion}
\label{sec:discuss}

\subsection{Formation of Low-Density Neptune-Size Planets}

\subsubsection{Core Nucleated Accretion}

Can core nucleated accretion form low-density planets in the size range of 2-6~$R_{\oplus}$? The answer is yes, given appropriate conditions. The solids surface density in the protoplanetary disk must be appropriate for the accretion of heavy element cores a few times as massive as Earth. These cores must grow early enough to accrete significant gaseous envelopes, but gas accretion must end early enough to avoid runaway gas accretion. Our evolution calculations in Section~\ref{sec:pform} demonstrate that $M_p<10~M_{\oplus}$ H/He rich planets can form for plausible choices of $\sigma$ and disk lifetimes. The values we chose for $\sigma$ are only slightly above that in the minimum-mass solar nebula, but high enough so that Jupiter at 5 AU can form in 4 Myr.

A second related question is whether core nucleated accretion with subsequent migration can lead to Neptune-size planets at high irradiation temperatures $T_{eq}\geq500$~K. Our evolution calculations uncover two factors that complicate achieving 2-6~$R_{\oplus}$ planets following inward migration. First, high irradiation temperatures lead to very large fluffy planets with radii $R_p>6~R_{\oplus}$. Second, very long migration timescales are required to heat a planet to $T_{eq}=1000~K$ while keeping its envelope intact. We elaborate both of these points below.

The salient feature of our evolution calculations is that, despite $M_p<10~M_{\oplus}$, the irradiated planet radii at 1 and 4~Gyr  are, in many cases, larger than $6~R_{\oplus}$. Specifically, all cases in Table~\ref{tab:Rp} with $T_{eq}=1000$~K or $M_p>5~M_{\oplus}$ have radii in excess of $6~R_{\oplus}$. Lower mass envelopes are required to yield Neptune-size planets at these high irradiation levels (Figure~\ref{fig:RParam}). Truncating gas accretion earlier (shorter disk lifetime) and subsequent envelope mass loss are two potential avenues toward $R_p<6~R_{\oplus}$. While the model radii at 1000~K are very uncertain due to uncertainties in the opacities near the radiative-convective boundary, for the cases in Table~\ref{tab:Rp} the conclusion that $R_p>6~R_{\oplus}$ is, nonetheless, robust. 

We found that slow planetary migration is needed for the low-mass envelopes to stay bound as the  temperature at the planetary surface increases. In our evolution calculations, the planets initially assemble at $T_{neb}=115$~K or 125~K and then migrate inward to $T_{eq}=500$~K or 1000~K. The long migration timescale ($\sim40$~Myr) taken to reach $T_{eq}=1000$~K with the envelope intact is in tension with typical disk lifetimes (1--10~Myr). The migration timescale to reach 500~K ($\sim5$~Myr) is more plausible. It is possible that evaporative cooling or increases in the envelope mean molecular weight from preferential loss of hydrogen could help the envelope remain bound. The planet evolution tracks presented do not include mass-loss. 

There do exist Neptune-size equilibrium configurations at $T_{eq}\geq500$~K for planets with H/He envelopes from core nucleated accretion. Our equilibrium planet structure models  in Section~\ref{sec:eqModel} explore and map out the ($M_{env}$, $M_{c}$, $T_{eq}$, $T_{int}$) parameter space that yield radii within the range 2--6~$R_{\oplus}$ (Figure~\ref{fig:RParam}). It is important to note, however, that the equilibrium models provide ``snap shots" of possible equilibrium configurations of planets undergoing quasi-static evolution. The models do not address how a planet could reach a given state, nor the timescale for the planet to evolve out of the state. 

\subsubsection{Outgassing of H$_2$}

The second formation pathway to low-density Neptune-size planets we considered was outgassing of H$_2$ from rocky planets. Outgassed low-mass planets $\left(M_p<30~M_{\oplus}\right)$ without substantial H$_2$O envelopes, however, can only account for radii up to $\sim3~R_{\oplus}$. Even achieving 3~$R_{\oplus}$ with an outgassed envelope is a stretch, requiring (concurrently) a near-optimal initial planetesimal composition, full oxidization of accreted metals, and retention of most H$_2$ released. Realistically, the majority of outgassed planets will be smaller than this radius limit, as we elaborate below. 

The metal and H$_2$O content of the primordial material forming a planet set a strict limit on the amount of H$_2$ that can be released via the outgassing reaction, 2Fe~+~3H$_2$O~$\rightarrow$~Fe$_2$O$_3$~+~3H$_2$. In this work, we have adopted a primordial chemical composition representative of EH chondrites mixed with just enough additional H$_2$O ice to fully oxidize all the metals. Out of the Solar System chondrites, this composition should be near optimum for outgassing of H$_2$ due to the high proportion of unoxidized iron (in metal or sulfide form) \citep{ElkinsTanton&Seager2008aApJ}. Typically planets forming from a mixture of Solar System chondrite-like material \citep[within which the proportion of metallic iron varies from 0.1 to 22 wt. \%]{Jarosewich1990Meteoritics} would have a lower capacity to outgas H$_2$. 

Even given a high initial amount of reduced metals, a planet's eventual outgassed envelope mass is contingent upon the fraction of metals that oxidizes. To bound the radii of outgassed planets, we considered the end-member case of complete oxidation of all Fe to Fe$_2$O$_3$. In this limiting case, the planet is core-less with all its iron incorporated in the mantle as oxides \citep{ElkinsTanton&Seager2008bApJ}. Planets retaining a metallic core would degas less H$_2$. Ultimately, the overall fraction of Fe that reacts with water is determined by the competition between the rate of oxidation and the rate of sinking of metallic Fe to form the planet iron core. For a more detailed discussion see \citet{ElkinsTanton&Seager2008bApJ}.

Finally, the mass-radius relations for outgassed planets in Section~\ref{sec:outgas} considered 100\% retention of all outgassed H$_2$. Atmospheric escape leads to less massive H$_2$ envelopes and smaller planets overall (Section~\ref{sec:mdot}). Indeed, while the primary outgassed atmospheres surrounding Earth and Mars during their accretion were likely H$_2$-dominated \citep{Schaefer&Fegley2010bIcarus}, both planets today harbor secondary atmospheres with higher mean molecular weights. 

How close can outgassed-planet radii plausibly get to the limiting outgassing $M_p-R_p$ relation? 
Relaxing our assumptions of optimum outgassing conditions, we investigate an intermediate, incomplete-oxidation case in which 50\% of the accreted Fe is converted to FeO. This scenario leads to planets with 19.5\% of their (initial) mass in an iron core, 0.6\% by mass degassed in H$_2$, and a mantle Mg\# of 0.62 (Table~\ref{tab:Outcomp}). 
With no loss of H$_2$ these planets could have radii up to 2.7~$R_{\oplus}$ (again considering $M_p\leq30~M_{\oplus}$). Whereas, with atmospheric mass loss, planets that retain only 1 to 10\% of the degassed H$_2$ would have radii up to at most 2.4--2.5~$R_{\oplus}$ for $T_{eq}=500-1000$~K. Thus, radii up to $\sim 2.5~R_{\oplus}$ are more realistically achieved by rocky planets with outgassed H$_2$ envelopes but no free water. Planets with a water layer between the rocky interior and H$_2$ envelope could be slightly larger, but only if little or no H$_2$ was mixed in with the H$_2$O.

\subsection{Maximum Planet Radius at Specified Mass}

We have modeled the internal structure of low-mass, large-radius planets with hydrogen-dominated atmospheres. For planets with outgassed H envelopes, we derived a limiting low density  $M_p-R_p$ relation by leveraging an upper bound on the amount of H$_2$ that can be degassed from rocky planetesimals. The limiting low-density $M_p-R_p$ relation is less clear cut for planets formed  from core nucleated accretion, because the initial reservoir of H/He accreted from the nebula need not be a constraining factor. Our detailed planet formation calculations provide discrete examples of planets at $T_{eq}=1000$~K with only a few Earth masses yet radii larger than Jupiter. 

The low-density limit for planets formed from core nucleated accretion depends on the heavy element core and envelope masses achievable at a given equilibrium temperature. The plausible combinations of ($M_{env}$, $M_{c}$, $T_{eq}$) in turn rely on the protoplanetary disk properties and the migration history of the planet. The heavy element core mass is determined by the isolation mass, given the solid surface density and the distance from the star where the planet forms. The isolation mass (and thus $M_c$) can have a wide range of values, from less than $0.1~M_{\oplus}$ to more than $20~M_{\oplus}$. The initial mass of H/He accreted by the planet is determined by the availability of a gas supply from the disk as governed by disk lifetime relative to the time taken for the heavy element core to reach isolation mass. Disk lifetimes range over an order of magnitude -- from 1 to 10~Myr, with a characteristic value of a few Myr \citep{Hillenbrand2008PhysScr} -- leading to some freedom in the initial $M_{env}$ expected from core nucleated accretion. Mass loss over the planet's history would serve to decrease $M_{env}$ over time. Finally, the current equilibrium temperature $T_{eq}$ depends on the migration history of the planet, and can, in principle, be anywhere from 100~K to 2000~K. Thus,  due to the large spread in observed disk properties, a wide range of ($M_{env}$, $M_{c}$, $T_{eq}$) from core nucleated accretion are plausible. We have shown detailed planet formation calculations for four reasonable choices of disk planetesimal densities and lifetimes.

We have succeeded in placing a tighter constraint on the low-density $M_p-R_p$ relation for outgassed planets than we have for planets from core nucleated accretion. This is due to the inherent limits on outgassed envelope masses; at very most, only a few percent of the mass of a planet can be outgassed in H$_2$. The end-member case of a planet that accreted from an optimum mixture of EH material and H$_2$O ice, where all the water and iron reacted, and where all released H$_2$ was retained, sets an upper bound on the transit radius possible at a given mass for a rocky planet with out-gassed H$_2$ atmosphere (Figure~\ref{fig:outgasR}). Typically, rocky planets with out-gassed H$_2$ atmospheres would have mean densities above this limiting $M_p-R_p$ relation. It should be noted, that our limiting $M_p-R_p$ relation applies to planets formed from material similar to Solar System chondrites. Planets formed from material with higher metallic Fe content would have the potential to outgas more H$_2$.

We have so far considered either core nucleated accretion or outgassing due to water-iron reactions as separate pathways for planets to acquire hydrogen rich envelopes. Core nucleated accretion contributes near solar composition material to the envelope ($Y\sim0.25$), while water-iron reactions contribute hydrogen but not helium ($Y=0$).  If both processes occur on the same exoplanet, an envelope with intermediate, sub-solar, non-zero Helium content  ($0<Y<0.25$) may result.

The assumed chemical make-up of the planet envelope and heavy element core affect the planet $M_p-R_p$ relations for planets formed by core nucleated accretion and by outgassing.
H/He envelopes in which He is depleted relative to solar will be more voluminous, for the same envelope masses, temperatures, and heavy element core properties. This is largely due to the influence of the mean molecular weight on the atmospheric scale height. For instance, decreasing Y=0.25 to Y=0.0 in the equilibrium planet models of Section~\ref{sec:eqres}, increases the radial extent of the envelopes by $\sim15-20\%$ for $M_p>20~M_{\oplus}$. 
For lower mass planets, the change in the gravitational acceleration between the top and bottom of the envelope can be substantial and $Y$ can have a larger effect on the envelope thickness. Pure H envelopes can be up to twice as thick as the corresponding H/He envelope, near the low mass extreme of the $M_p-R_p$ relations in Section~\ref{sec:eqres}. In our planet structure models, however, the effect of the envelope He abundance is largely offset by the higher density heavy element core composition in our outgassing models (EH chondrite cores) as compared to our core nucleated accretion models (ice-rock cores). 

We have mapped out the contribution of low-mass planets with hydrogen-dominated atmospheres to the limiting low-density $M_p\left(R_p\right)$ relation. 
Although we have not considered them in detail here, planets may also form with high molecular weight envelopes, for instance, after having accreted large amounts of ices beyond the snow line  \citep[e.g., ][]{Kuchner2003ApJ, LegerEt2004Icarus}. Higher molecular weight envelopes are more dense (with smaller atmospheric scale heights) than their hydrogen-dominated counterparts, but may be less affected by atmospheric escape. 
It is possible that planets with high molecular weight atmospheres could also contribute to the limiting low-density $M_p\left(R_p\right)$ relation for Neptune-size planets. 

\subsection{Minimum Planet Mass at Specified Radius}
\label{sec:mmin}

Our ideal goal was to determine a lower bound on the plausible planet mass given a planet radius in the range 2 -- 6~$R_{\oplus}$ and equilibrium temperature $T\geq500$~K. We note that the relation defining the the maximum radius for a given planet mass does not necessarily translate into a relation for the minimum planet mass at a given radius. Indeed, at low masses, $dR_p/dM_p<0$ in the iso-composition $M_p-R_p$ relations for planets with gas envelopes (e.g., Figures~\ref{fig:MRgmf}, \ref{fig:MRTint} and \ref{fig:outgasR}). Thus, in order to bracket the minimum planet mass of a transiting planet candidate, we must assess the survivability of low-mass planets for a range of interior compositions. 

Mass loss is a major limiting factor that constrains the minimum $M_p\left(R_p\right)$ for strongly irradiated ($T\geq500$~K) Neptune-size planets harboring hydrogen dominated envelopes (Section~\ref{sec:mdot}). This is true whether the planet acquired its envelope through core nucleated accretion or through outgassing. If the heavy element core mass is small ($\lesssim2~M_{\oplus}$) and $T_{eq}$ is high (1000 K) then the planet will not be able to hold on to very much gas. With the energy limited mass loss rates from Equation~(\ref{eq:mdot}), we may roughly assess the survivability of potential planet configurations.  By choosing a threshold envelope loss timescale $t_{\dot{M}0}$, we can derive a lower bound on the planet mass at a given radius based on the requirement $t_{\dot{M}}\geq t_{\dot{M}0}$. To illustrate this approach, we adopt $t_{\dot{M}0}=1$~Gyr and explore what this implies for planets with low mean molecular weight envelopes from core nucleated accretion (Figure~\ref{fig:RParam}) and from outgassing (Figure~\ref{fig:outRParam}).

We estimate, using Figure~\ref{fig:RParam}, the minimum masses of $R_p=2$--$6~R_{\oplus}$ planets with H/He envelopes formed by core nucleated accretion beyond the snow line. For $R_p=6~R_{\oplus}$ planets, the least massive planet models that satisfy $t_{\dot{M}}\geq1$~Gyr are 1.3 to $1.7~M_{\oplus}$ at $T_{eq}=500$~K, and 4.0 to $4.7~M_{\oplus}$ at $T_{eq}=1000$~K (for $L_{p}/M_p$ between $10^{-9}$ and $10^{-11}~\rm{W\,kg^{-1}}$).
Analogously, at $R_p=4~R_{\oplus}$, the $t_{\dot{M}}\geq1$~Gyr survivability constraint requires that $M_p\gtrsim 1.1$ to $1.4~M_{\oplus}$ at $T_{eq}=500$~K, and $M_p\gtrsim 3.6$ to $4.3~M_{\oplus}$ at $T_{eq}=1000$~K. 
At $R_p=2~R_{\oplus}$, almost all possible ($M_p$, $M_{env}$, $L_p$, $T_{eq}$) configurations in Figure~\ref{fig:RParam} have sub-gigayear envelope-loss timescales, due to the small planet and envelope masses ($M_p<5~M_{\oplus}$, $0\leq M_{env}< 0.1\%~M_p$). An ice/rock core surrounded by an H/He envelope from core nucleated accretion may not be a plausible interior composition scenario for $2~R_{\oplus}$ planets at these equilibrium temperatures. Instead, other possibilities not considered here (e.g., high molecular weight envelopes, or envelope-less planets) may account for the minimum plausible planet mass at $2~R_{\oplus}$. 

We turn now to planets with outgassed hydrogen envelopes but no surface water, and apply the envelope mass loss threshold to  Figure~\ref{fig:outRParam}. In addition to atmospheric escape, hydrogen-rich envelopes acquired by outgassing are also constrained by the limited H$_2$ reservoir (magenta line in Figure~\ref{fig:outRParam}). We find that, at $T_{eq}=500$~K, there exist potential planet configurations that satisfy $t_{\dot{M}}\geq 1$~Gyr with masses as low as $1~M_{\oplus}$ for planet radii ranging from 2 to 3~$R_{\oplus}$. Granted, these minimum-mass outgassing scenarios necessitate near maximal release and retention of H$_2$. In contrast, at $T_{eq}=1000$~K, all possible $H_2$ envelopes leading to $R_p=2~R_{\oplus}$ have sub-gigayear envelope-loss timescales. For larger radii (2.5, and $3~R_{\oplus}$), planets with masses as low as 3.5 -- 4~$M_{\oplus}$ (depending on $L_p$) may pass the $t_{\dot{M}}\geq1$~Gyr survivability criterion.

We emphasize that minimum masses estimated following the approach above are contingent upon the chosen $t_{\dot{M}}$ threshold, the energy-limited mass loss parameter values assumed (here we took $\epsilon L_{XUV}/L_{BOL}=10^{-6}$), and the range of planet ages/intrinsic luminosities under consideration. High (lower) $t_{\dot{M}0}$ would leach to higher (lower) minimum $M_p(R_p)$. Although quantitatively very assumption-dependent, minimum masses derived from $t_{\dot{M}0}$ may nonetheless yield important qualitative insights.

\subsection{Implications for  {\it Kepler} Planet Candidates}

We conclude with a discussion of the implications of our results for the Neptune-size planet candidates discovered by {\it Kepler}. Candidates in the  $2-6~R_{\oplus}$ size-range account for a large fraction of the current candidates detected by  {\it Kepler} \citep{BoruckiEt2011ApJ, BoruckiEt2011AstroPh}. This raises the question of why Neptune-size planet candidates are so common. One possible contributing factor revealed by this study is that not very much mass is needed in a hydrogen dominated envelope for a rocky heavy element core to reach radii within 2-6~$R_{\oplus}$.

Our main conclusion is that the Neptune-size planet candidates could have low mass $\left(M_p<4~M_{\oplus}\right)$. This deduction is supported by our calculations of the formation, structure, and survival of planets with voluminous envelopes of light gasses. 

\begin{itemize}
\item {\bf Formation:} We demonstrated that planets 3 to 8~$M_{\oplus}$ with substantial H/He envelopes can plausibly form by core nucleated accretion beyond the snow line and migrate to $T_{eq}\sim500$~K given reasonable disk surface densities and disk dissipation timescales. Migration to $T_{eq}\sim1000$~K with the envelope intact in timescales of a few Myr is more challenging. 
\item  {\bf Structure:} We mapped the regions of ($M_p$, $M_{env}$, $T_{eq}$, $L_p$) parameter space that yield radii between 2 and $6~R_{\oplus}$ for planets with H/He envelopes from core nucleated acretion and for planets with outgassed H$_2$ envelopes (Figures~\ref{fig:RParam} and \ref{fig:outRParam}, respectively).  Since at most a few percent of a planet's mass can be degassed as H$_2$, rocky super-Earths ($M_p<30~M_{\oplus}$) with outgassed hydrogen atmospheres but without substantial H$_2$O typically will not account for {\it Kepler} planet candidates larger than $\sim3~R_{\oplus}$. 
\item {\bf Survival:} Envelope mass loss plays a major role governing the minimum plausible $M_p\left(R_p\right)$ for strongly irradiated ($T\geq500$~K) Neptune-size planets with hydrogen-dominated envelopes.  At $R_p=2~R_{\oplus}$, H/He envelopes surrounding ice-rock cores would likely be lost in short order. At larger radii (2.5 to 6~$R_{\oplus}$), planet configurations with envelope mass loss timescales longer than a Gyr (assuming $\epsilon L_{XUV}/L_{BOL}=10^{-6}$) exist down to masses $\sim1~M_{\oplus}$ at $T_{eq}=500$~K and down to $\sim4~M_{\oplus}$ at $T_{eq}=1000$~K.
\end{itemize}
 
Neptune-size planets with masses $M_p<4~M_{\oplus}$ could prove a challenge for radial velocity (RV) follow-up due to their low RV semi-amplitudes, but confirmation and mass measurements through transit timing variations may be possible in some cases \citep[e.g., Kepler-11,][]{LissauerEt2011Nature}. Figures~\ref{fig:RParam} and \ref{fig:outRParam} may be useful tools for assessing minimum masses for {\it Kepler} planet candidates.

\acknowledgments

P.B. and J.L acknowledge support from the NASA origins program. P. B. 
acknowledges support from NSF grant AST0908807. The authors thank Olenka Hubickyj for making Figures 1, 2, and 3. We also thank Linda Elkins-Tanton for helpful discussions about planet outgassing and silicate mantles, Benjamin Weiss for his insights into EH chondrite compositions, and Geoff Marcy and an anonymous referee for valuable comments on the manuscript. 
\;

\;

\bibliography{exoplanets}

\clearpage

\end{document}